\documentstyle[sprocl,epsfig]{article}

\def\Journal#1#2#3#4{{#1} {\bf #2}, #3 (#4)}

\def\AP{\em Ann. Phys.}

\def\NPA{{\em Nucl. Phys.} A}
\def\NPB{{\em Nucl. Phys.} B}
\def\PLB{{\em Phys. Lett.}  B}
\def\PRL{\em Phys. Rev. Lett.}
\def\PRC{{\em Phys. Rev.} C}
\def\PRD{{\em Phys. Rev.} D}
\def\ZPC{{\em Z. Phys.} C}

\def\be{\begin{equation}}
\def\ee{\end{equation}}
\def\bea{\begin{eqnarray}}
\def\eea{\end{eqnarray}}

\newcommand{\case}[2]{\mbox{\footnotesize $\displaystyle \frac{#1}{#2}$}}
\newcommand{\lsim}{\mathrel{\rlap{\lower3pt\hbox{\hskip0pt$\sim$}}
\raise2pt\hbox{$<$}}}
\newcommand{\gsim}{\mathrel{\rlap{\lower3pt\hbox{\hskip0pt$\sim$}}
\raise2pt\hbox{$>$}}}

\begin{document}

\title{NONPERTURBATIVE QCD WITH MODERN TOOLS} 

\author{C. D. ROBERTS}

\address{Physics Division, Bldg. 203, Argonne National Laboratory\\
Argonne IL 60439-4843, USA}


\maketitle\abstracts{In these lectures I introduce and explore a range of
topics of contemporary interest in hadronic physics: from what drives the
formation of a nonzero quark condensate to the effect that mechanism has on
light and heavy meson form factors, and the properties of the quark-gluon
plasma.  The trail leads naturally through a discussion of confinement,
dynamical chiral symmetry breaking and bound state structure: phenomena that
require nonperturbative methods for their explanation.  In all of this, the
necessary momentum-dependent modification of the quark and gluon propagators
plays a significant role.}

\section{Hadron Physics}
\label{intro}
The strong interaction spectrum has a regularity that can be understood
simply.  Mesons are composed of a constituent quark and its antiparticle, and
baryons of three constituent quarks.  The quarks have quantum numbers: colour
- red, blue, green; flavour - up, down, strange, etc., and spin - up or down.
Attributing a flavour-dependent mass to these constituent quarks: $M_{u}
\approx M_d \sim 320\,$MeV, $M_s \sim 400\,$MeV, and demanding that only
colour singlets exist, one obtains a quantitative description that includes
the multiplet structure.  Almost.

In this picture the pion is a pseudoscalar meson with the spins of its
constituents antiparallel: $m_\pi \approx 140\,$MeV.  Align the spins and one
has a vector meson, which is identified with the $\rho$: $m_\rho \approx
770\,$MeV.  The simplest explanation of the mass ratio in
Eq.~(\ref{dcsb})$\,$\cite{pdg96} 
\begin{equation}
\label{dcsb}
\begin{array}{lc|clc|cl}
\displaystyle\frac{m_\rho^2}{m_\pi^2} = 30 & & &
\displaystyle\frac{m_{\pi^\prime}^2}{m_\pi^2} = 86 & & &
\displaystyle\frac{m_N^2}{m_\pi^2} = 45  \\
&&&&&&\\[-0.7em]
\displaystyle \frac{m_{a_1}^2}{m_{a_0}^2} = 1.7 & & &
\displaystyle\frac{m_{\rho^\prime}^2}{m_{\rho}^2} = 3.5 & & &
\displaystyle\frac{m_N^2}{m_\rho^2} = 1.5
\end{array}
\end{equation}
is hyperfine splitting, akin to the splitting of each level in the hydrogen
atom that measures whether the spin of the electron and proton are aligned or
not.  A difficulty with this interpretation, however, is that the
pseudovector- to scalar-meson mass ratio, $m_{a_1}^2/m_{a_0}^2$, should have
the same origin but it is much smaller.  Equation~(1) also shows that the
excitation energy in the pseudoscalar channel, which is measured by the mass
difference between the pion and its first excited state, is much larger than
that in the vector channel.  As a final illustration of the peculiarity of
the pion, I note that the expected value of the ratio in the third column is
$\approx 2$ because it should just count the number of constituents in a
baryon as compared to a meson.  It is for the $\rho$ but not for the $\pi$.
The ``unnaturally'' low mass of the $\pi$ has many consequences; e.g., it
means that attraction in the $N$-$N$ interaction persists over a much longer
range than repulsion, which underlies the line of nuclear stability and makes
possible the existence of heavy elements.

Although the $\pi$ was the first meson discovered, its peculiar properties
are still not understood widely.  They are almost completely determined by
the phenomena driving dynamical chiral symmetry breaking [DCSB].  That is an
intrinsically nonperturbative effect, the understanding of which is one focus
of these lecture notes.

Quarks provide a means of understanding much of the regularity in the hadron
spectrum.  However, they do not form part of that spectrum: quarks carry the
colour quantum number and the spectrum consists only of colourless objects.
Are they then a mathematical artifice useful only as a means of realising
group theory?  No, they are observed in inclusive, deep inelastic scattering:
$e\,p \to e^\prime\, +\,$``debris'', even though the debris never contains an
isolated quark.

The cross section for deep inelastic scattering from a proton target is
\begin{equation}
\displaystyle
        \frac{d^2\sigma}{dQ^2\,d\nu} =
        \frac{\alpha^2}{4 E_e^2 \sin^4\frac{\theta}{2}}
        \left(2 W_1(Q^2,\nu)\,\sin^2\frac{\theta}{2}
                        + W_2(Q^2,\nu)\,\cos^2\frac{\theta}{2}\right)\,,
\end{equation}
where $\theta$ is the scattering angle, $Q^2$ is the spacelike squared
momentum transfer, and $\nu= E_{e^\prime} - E_{e}$ for $\vec{p}_p=0$.  It is
characterised by two scalar functions: $W_1$, $W_2$, which contain a great
deal of information about proton structure.  If the proton is a composite
particle composed of pointlike constituents, then it follows that for $Q^2
\to \infty\,, \nu\to\infty\,$ with $\nu\gg Q^2 $
\begin{equation}
\label{disW}
\begin{array}{ccc}
        W_1(Q^2,\nu^2)& \to& F_1({ Q^2/\nu^2})\,,\\
        \nu W_2(Q^2,\nu^2)&\to& F_2({ Q^2/\nu^2})\,;
        \end{array}
\end{equation}
i.e., that in inclusive, deep inelastic processes $W_{1/2}$ are functions
only of $x:= Q^2/\nu^2$ and not of $Q^2$ and $\nu^2$ independently.  

Such experiments were first performed at SLAC in the late sixties and have
been extended widely since then.  Equation~(\ref{disW}) is confirmed
completely,\cite{pdg96} it is correct as $Q^2$ varies over four
orders-of-magnitude.  The presence of pointlike constituents is the only
possible explanation of this behaviour, and many other observations made in
such experiments.  Hence quarks exist but are ``confined'' in colourless
bound states, and understanding the origin and meaning of that confinement is
another focus of these notes.

The role that quarks play in hadron structure is a primary subject of
contemporary experimental and theoretical nuclear physics and it is not
necessary to employ high energy or large momentum transfer to explore it.
The electromagnetic form factor of the neutron is a good example.  The
differential cross section for electron-neutron scattering:
\begin{equation}
\frac{d\sigma}{d\Omega} = \left. \frac{d\sigma}{d Q^2}\right|_{\rm Mott}
\left(\frac{ G_E^2(Q^2) + \tau\,G_M^2(Q^2)}{1 + \tau} + 2 \tau
\,G_M^2(Q^2)\tan^2\frac{\theta}{2} \right)\,,
\end{equation}
$\tau:= Q^2/(4 M_N^2)$, involves the ``Mott''-term, which describes electron
scattering from a point source, and the electric and magnetic form factors,
$G_E$ and $G_M$, that describe the structure of the neutron.  Of particular
interest is the electric form factor.  For a neutral particle $G_E(Q^2=0)=0$,
and it is identically zero for a neutral point particle.  Therefore
deviations from zero are a good measure of ones description of the internal
structure of the neutron.

The neutron charge radius has been measured$\,$\cite{ncharge} in the
transmission of low energy neutrons through Pb atoms:
\begin{equation}
r^2_n:= \,-6\,
\left. \frac{d G_E(Q^2)}{d Q^2} \right|_{Q^2=0}
= \, - 0.113 \pm 0.003 \,{\rm fm}^2\,.
\end{equation}
It is negative and that must be understood.  In
studies$\,$\cite{brt96,fredmike} of $K^0$ and $K^{0\ast}$ mesons one finds
$r^2_{K^0} < 0 $ and $r^2_{K^{0\ast}} < 0 $.  These results are understood
simply: the mesons are $\bar s$-$d$ bound states and, being lighter, the
negatively charged $d$-quark is able to propagate further from the system's
centre-of-mass thereby generating a negative charge-distribution at
long-range.  That mechanism generates 60\% of the observed
value:$\,$\cite{molzon} $r^2_{K^0}= -0.054 \pm 0.026\,$fm$^2$, and something
similar is likely to be responsible for a large part of $r^2_n$.

\begin{figure}[t]
\centering{ \epsfig{figure=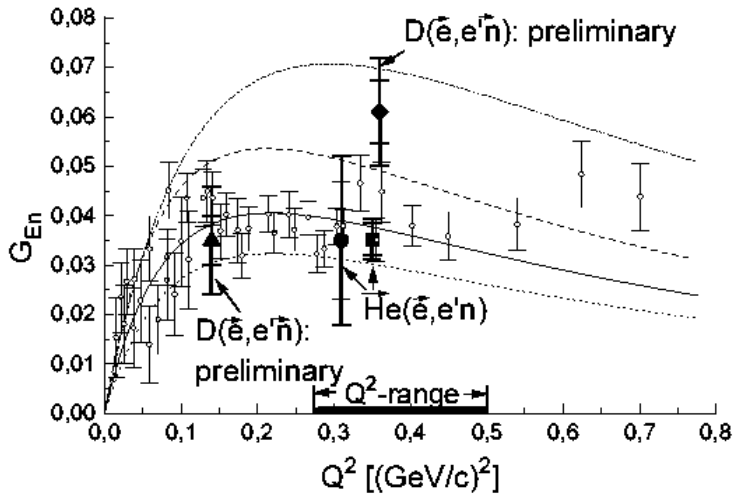,height=7.0cm} }
\caption{$G_E^n$ - summary of data.\protect\cite{drechsel}.  The lines are
two-parameter fits using different $N$-$N$ potentials in the calculation of
the deuterium wave function.  The ``$Q^2$-range'' is that of a recent MAMI
[Mainz] experiment: $\!^3 \vec{\rm He}(\vec{e},e^\prime n)$.
\label{neutron}}
\end{figure}

The $Q^2$-evolution of the form factors is more difficult to measure.  In the
absence of free neutron targets it must be inferred from the scattering of
electrons from loosely-bound collections of polarised nucleons; e.g.,
deuterium or $\!^3$He.  However, even in such loosely bound systems the
inferred cross section is sensitive to details of the model used to calculate
the nuclear wave function and the current operator, although it is
anticipated that these effects can be minimised in double polarisation
experiments.\cite{drechsel} The present status of measurements of $G_E^n$ is
illustrated in Fig.~\ref{neutron}.

Such complications often arise in prospecting for quarks and gluons.
However, their role in hadron structure is identifiable only in the
signatures they write in, for example, the $Q^2$-dependence of form factors,
so the difficulties must be overcome.  The accurate calculation of form
factors on the entire domain of accessible $Q^2$ can be a very good test of
any given description of hadron structure.  It relies on combining an
understanding of confinement and DCSB with that of the intrinsically
nonperturbative analysis of bound state structure.  This synthesis is the
primary focus of these notes.

Given that at low-energy and small-$Q^2$ the strong interaction is
characterised by the nonperturbative phenomena of confinement, DCSB and the
behaviour of colourless bound states, is its study ever simple?  Yes, in
scattering processes at high-energy or large-$Q^2$ where perturbation theory
is applicable because QCD is asymptotically free.  As described in
Sec.~\ref{pQCD}, the coupling constant evolves with $Q^2$:
\begin{equation}
\alpha_{\rm S}(Q^2) \stackrel{Q^2\to \infty}{\longrightarrow} 0\,,
\end{equation}
and hence quarks and gluons behave as weakly interacting, massless particles
when the momentum transfer is large.

This also leads to the identification of another nonperturbative phenomenon.
If one introduces the intensive variables: temperature, $T$, and quark
chemical potential, $\mu$, QCD acquires additional mass-scales with which the
coupling can {\it run}.  Hence,
\begin{equation}
\alpha_{\rm S}(Q^2=0;T ,\mu )\sim 0\,\; {\rm for}\;
T\gg \Lambda_{\rm QCD} \;{\rm  and/or}\; \mu\gg \Lambda_{\rm QCD}\,,
\end{equation}
where $\Lambda_{\rm QCD}\sim 200\,$MeV is the intrinsic,
renormalisation-induced mass-scale in QCD.  It follows that at
finite-$(T,\mu)$ there is a phase of QCD in which quarks and gluons are
weakly interacting, {\em irrespective} of the momentum
transfer;$\,$\cite{collinsperry} i.e., a quark-gluon plasma [QGP].  In this
phase confinement and DCSB are absent and the nature of the strong
interaction spectrum is qualitatively different.  Phase transitions are
intrinsically nonperturbative phenomena.

\subsection{Modern Experimental Tools}
\label{subsec:MET}
Understanding the strong interaction, applying nonperturbative methods in
QCD, is a difficult problem.  As in all of physics, progress here and the
development of understanding is promoted by the challenge of explaining real
data, and there is a new generation of experimental facilities whose goal is
to furnish this challenge.

DESY$\,$\cite{desy} [Deutsches Elektronen-Synchroton] in Hamburg is the site
of the Hadron-Electron Ring Accelerator facility [HERA].  It is a collider:
in a $6\,336\,$m underground tunnel a longitudinally-polarised, $30\,$GeV
$e^-$ or $e^+$ beam collides with a counter-circulating, $820\,$GeV proton
beam at four interaction zones.  There are currently three large experimental
collaborations using HERA: H1 [since 1992], ZEUS [since 1992] and HERMES
[since 1995], with the goal of elucidating the internal structure of the
nucleon.

Newport News, Virginia, is the site of the Thomas Jefferson National
Accelerator Facility$\,$\cite{tjnaf} [TJNAF, formerly CEBAF].  This machine
currently accelerates electrons to $4\,$GeV in five circuits of an
oval-shaped, $1\,400$m underground tunnel.  The electrons are then focused
into three large, underground target areas: Halls - A [$50\,$m diameter], B
[$30\,$m] and C [$50\,$m].  Although Hall B is not yet fully operational,
experiments have been underway in Halls A and C for two years.  Many have
been completed and other important experiments are underway; e.g.,
Hall C is currently hosting a measurement of the electromagnetic pion form
factor, which will probe the evolution from the nonperturbative to the
perturbative domain in QCD, and that of $G_E^n$ is scheduled.  The facility
plans to expand the research capability of the accelerator by raising the
beam energy to $12\,$GeV by the year 2006.

Cornell University in Ithaca, New York, is the site of the Cornell Electron
Storage Ring$\,$\cite{cesr} [CESR].  A circular $e^+$-$e^-$ collider, $12\,$m
underground with a circumference of $768\,$m, CESR is a $b$-quark factory.
With the CESR collision energy tuned to that of the $\Upsilon(4s)$ $b$-$\bar
b$-resonance, which is massive enough to decay to a $B$-$\bar B$ meson pair,
the CLEO detector is able to observe the decays of $B$-mesons.  A careful
analysis of these decays can provide tight constraints on the elements of the
Cabibbo-Kobayashi-Maskawa [CKM] matrix, which describes the mixing of the
eigenstates of the strong and weak interactions.  Most of what is currently
known about $B$-meson decays is due to experiments using the CLEO detector.
However, since experiments using $B$-mesons are a far-reaching tool to probe
the standard model, it is anticipated that the next five years will see the
construction of three more $b$-factories: at KEK in Japan; at SLAC; and in
the completion of the HERA-B site at DESY.  Heavy-quark physics is a
burgeoning, contemporary focus.

Brookhaven National Laboratory$\,$\cite{bnl} [BNL] is the site of the AGS
[Alternating Gradient Synchrotron] and the future site of RHIC, the
Relativistic Heavy Ion Collider.  Its relativistic heavy ion programme
currently focuses on the study of nuclear matter at extremes of temperature
and density, and prospecting for the quark gluon plasma.  The AGS has been
operating since 1962 and its fixed target experiments are laying the
foundations of the search for the QGP.  In the future, the AGS will act as an
injector for RHIC, which is due for completion in 1999.  Using two concentric
rings, each $3.8\,$km in diameter, RHIC will collide two 100$\,$A$\,$GeV
$^{197}\!$Au beams at several detector sites around the ring.  Producing a
total centre-of-mass energy of $\sim 40\,$TeV at each collision site, RHIC is
expected to produce an equilibrated QGP, approaching it along a high-$T$,
baryon-poor trajectory.

Also in the hunt for the QGP is the SpS [Super-proton Synchrotron] at
CERN,\cite{cern} seven kilometres in circumference.  Fixed target experiments
with S$+$Pb at $\sim$ 200 A GeV and Pb $+$ Pb at $\sim\,$160$\,$A$\,$GeV are
possible using this machine, producing much higher centre-of-mass energies
than those available at the AGS.  With these resources, experiments
CERES/NA45 and NA50 have observed interesting effects: an enhancement in the
dilepton spectrum and a suppression of the $\psi$-production cross section.
Whether these signal the formation of a QGP is an unresolved issue but their
explanation is certainly providing a challenge to theory.  The next phase is
the construction of the LHC [Large Hadron Collider].  Currently anticipated
to be completed in 2005, it will share the $27\,$km LEP tunnel and be capable
of hosting heavy ion collisions that generate a centre-of-mass energy
thirty-times greater than that possible at RHIC.  If the QGP is not
discovered with RHIC then the LHC can still promise to do so.  If it is, then
the LHC should provide a tool for thoroughly exploring its properties, and
for many other things.

I have surveyed only a few of the many existing and planned experimental
facilities.  Nevertheless, the intensity of effort being expended in
prospecting for quarks and gluons, and in elucidating the subtle properties
of QCD, is obvious in this.  More detailed information is available at the
web sites I have indicated and obvious searches will yield details about
other facilities.
\subsection{Perturbative QCD}
\label{pQCD}
QCD is a gauge theory, which in some respects is like QED.  The elementary
fermions, quarks in this case, interact by exchanging massless vector bosons:
gluons.  In QCD the coupling is through the colour quantum number: the colour
charge is the analogue of the electric charge.  The special feature of QCD is
that the gluon carries the colour charge, as well as the quarks, and that
means the gluons are self-interacting, which is enough to completely change
the qualitative character of the theory.

\begin{figure}[t]
\vspace*{-1.0em}

\centering{\ \epsfig{figure=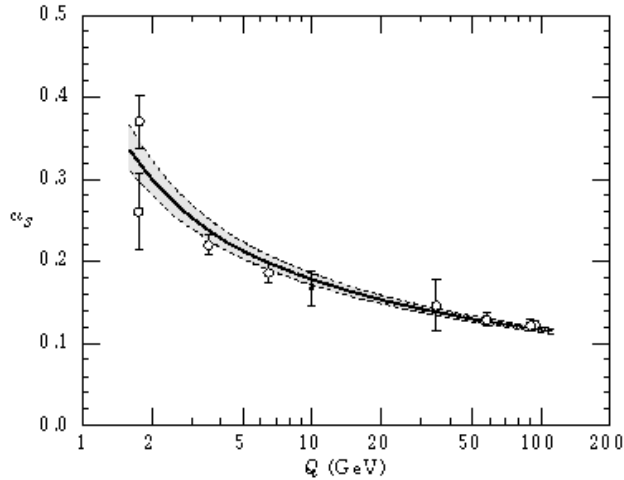,height=6.6cm}}
\caption{Summary$\,$\protect\cite{pdg96} of the values of $\alpha_{\rm
QCD}(Q)$ at the points where they are measured.  The lines show the central
values and the $\pm \sigma$ limits of the average.
\label{alphas}}
\end{figure}

In both QED and QCD the coupling ``constant'' is not constant at all but
evolves with $Q^2$ and it can be calculated perturbatively:
\begin{eqnarray}
\label{alphaQED}
\displaystyle\alpha_{\rm QED} & = & \frac{\alpha} {1 - \alpha/3\pi
        \ln\left(Q^2/m_e^2\right)}\,,\;\alpha\simeq \frac{1}{137}\,, \\[0.5em]
\label{alphaQCD}
        \alpha_{\rm S} & = & \frac{12\pi} {\left({ 11 N_c} - { 2
        N_f}\right) \ln\left(Q^2/\Lambda_{\rm QCD}^2\right)},
\end{eqnarray}
where the ``$11 N_c$'', $N_c=3$ is the number of colours, appears because the
gluons are self-interacting.  $N_f$ is the number of fermion flavours.  As
depicted in Fig.~\ref{alphas}, 
$\alpha_{\rm S}(Q)$ decreases with increasing-$Q$, which is opposite to the
behaviour of $\alpha_{\rm QED}(Q)$ quoted in Eq.~(\ref{alphaQED}).  The rate
of the evolution is also very different, being determined by vastly different
mass-scales: $\Lambda_{\rm QCD}/m_e \sim 450$.  The reduction in $\alpha_{\rm
S}$ with increasing $Q$ is called ``asymptotic freedom''.  It is due to gluon
self-interactions and entails that perturbation theory is valid at
large-$Q^2$ in QCD; i.e., at short-distances, {\it not} at long-distances as
in QED.

A very successful application of perturbative QCD is the calculation of
\begin{equation}
R := \frac{\sigma_{e^+ e^- \to {\rm hadrons}}}{\sigma_{e^+ e^- \to \mu^+
\mu^-}}\,,
\end{equation}
which is known to order $\alpha_{\rm S}^3$:
\begin{equation}
R = N_c\,\sum_{f=1}^{N_f} e_f^2 \,
\left[ 1 + \frac{\alpha_S}{\pi} + 1.411 \left(\frac{\alpha_S}{\pi}\right)^2
- 12.8 \left(\frac{\alpha_S}{\pi}\right)^3 + \ldots \right]\,,
\end{equation}
where $e_f$ is the electric charge of a quark of flavour $f$.\footnote{This
expression assumes that all the quarks are massless but it is also known for
massive quarks.  In comparing with data, $b$-quark mass corrections are
important.}  
\begin{figure}[t] 
\vspace*{-1.0em}

\centering{\ 
\epsfig{figure=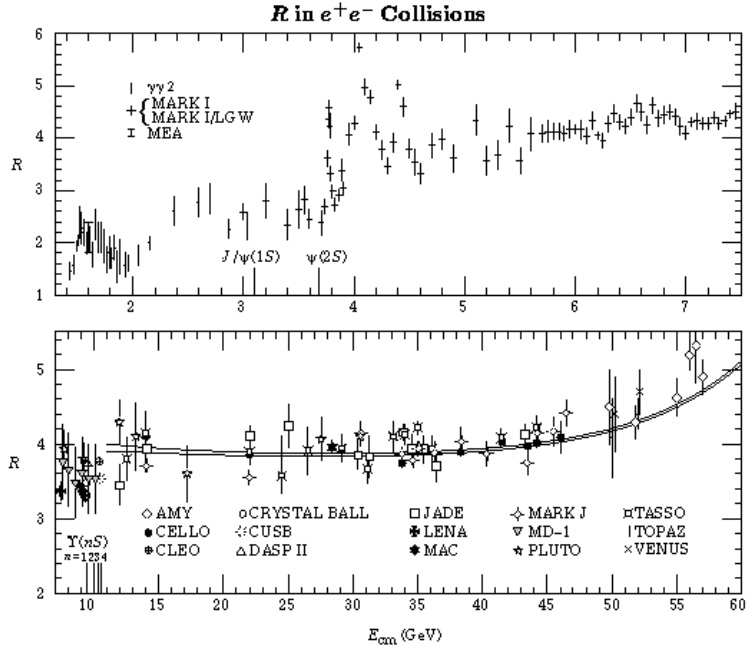,height=9.0cm}}
\caption{Selected measurements of $R$.\protect\cite{pdg96} The positions of
the $\psi$, $\psi(2S)$ and four lowest $\Upsilon(nS)$ resonances are also
indicated, which are not predicted by perturbation theory.  A fit to the data
in the lower panel yields $\alpha_{\rm S}(Q=34\,{\rm GeV})=0.142 \pm 0.03$.
The experimental error is larger than the theoretical uncertainty.  The
plotted calculations include electroweak corrections, which predict the
increase as $E_{cm} \to M_Z$.
\label{ratioR}}
\end{figure}
The theoretical prediction is compared with experiment in Fig.~\ref{ratioR}.
It proves that $N_c=3$ because the normalisation of the prediction is
otherwise incorrect.  The figure also highlights the current-quark mass
thresholds and confirms the electric charge assignments: below the $c$-quark
threshold $R \approx N_c\,\sum_{f=1}^{N_f= 3} e_f^2 = 2$, then $R \approx
N_c\,\sum_{f=1}^{N_f= 4} e_f^2 = 10/3$, and above the $b$-quark threshold $R
\approx N_c\,\sum_{f=1}^{N_f= 5} e_f^2 = 11/3$.

Perturbation theory is the most widely used, systematic tool in physics and
in QCD it is extremely accurate in high-energy processes.  In addition to the
calculation of $R$, there are many other applications which confirm QCD as
the theory describing the strong interaction.  However, there are domains and
problems in QCD that perturbation theory simply cannot describe.  This is
very obvious in Fig.~\ref{alphas} where the coupling is seen to {\it
increase} with the inter-particle separation: QCD becomes a strong coupling
theory for $Q<2\,$GeV [$ x > 0.1\,$fm].  Further, as observed in
Sec.~\ref{intro}, confinement, DCSB, bound state structure and phase
transitions are intrinsically nonperturbative phenomena.  They are at the
core of hadron physics and hence nonperturbative methods are essential in
prospecting for quarks and gluons.
\subsection{Nonperturbative Effects in QCD}\label{subsec:AC}
The extraordinary effects in QCD can be explained in terms of the properties
of {\it dressed}-quark and -gluon propagators.  They describe the
``in-medium'' propagation characteristics of QCD's elementary quanta, with
the ``medium'' being the nontrivial ground state of QCD.  A photon
propagating through a dense $e^-$-gas provides a familiar example of the
effect a medium has on the propagation of an elementary particle.  Due to
particle-hole excitations the propagation of the photon is modified:
\begin{equation}
\frac{1}{Q^2} \to \frac{1}{Q^2 + m_D^2}\,;
\end{equation}
i.e., the photon acquires an effective mass.  The ``Debye'' mass, $m_D
\propto k_F$, the Fermi momentum, and it screens the interaction so that in
the dense $e^-$-gas the Coulomb interaction has a finite range: $r\propto
1/m_D$.  Quark and gluon propagators are modified in a similar way.  They
acquire {\it momentum-dependent} effective masses, which have observable
effects on hadron properties.
\vspace*{0.5em}

\hspace*{-\parindent}{\it \arabic{section}.\arabic{subsection}.1}~{\bf Chiral
symmetry}.~Gauge theories with massless fermions have a chiral symmetry.  Its
effect can be visualised by considering the helicity: $\lambda \propto J\cdot
p$, the projection of the fermion's spin onto its direction of motion.
$\lambda$ is a Poincar\'e invariant spin observable that takes a value of
$\pm 1$.  The chirality operator can be realised as a $4\times 4$-matrix,
$\gamma_5$, and a chiral transformation is then represented as a rotation of
the $4\times 1$-matrix quark spinor field
\begin{equation}
q(x) \to {\rm e}^{i \gamma_5 \theta}\,q(x)\,.
\end{equation}
A chiral rotation through $\theta = \pi/2$ has no effect on a $\lambda = +1$
quark, $q_{\lambda =\, +} \to q_{\lambda =\, +}$, but changes the sign of a
$\lambda = -1$ quark field, $q_{\lambda =\, -} \to \,-\,q_{\lambda =\, -}$.
In composite particles this is manifest as a flip in their parity: $J^{P=\,+}
\leftrightarrow J^{P=\,-}$; i.e., a $\theta = \pi/2$ chiral rotation is
equivalent to a parity transformation.  Exact chiral symmetry therefore
entails that degenerate parity multiplets must be present in the spectrum of
the theory.

For many reasons, the masses of the $u$- and $d$-quarks are expected to be
very small; i.e., $m_u \sim m_d \ll \Lambda_{\rm QCD}$.  Therefore chiral
symmetry should only be weakly broken, with the strong interaction spectrum
exhibiting nearly degenerate parity partners.  The experimental comparison is
presented in Eq.~(\ref{ohdear}): 
\begin{equation}
\label{ohdear}
\begin{array}{lc|clc|cl}
N(\frac{1}{2}^+,938) &&& \pi(0^-,140) &&&\rho(1^-,770)\\ 
N(\frac{1}{2}^-,1535) &&& a_0(0^+,980) &&& a_1(1^+,1260)
\end{array}\,.
\end{equation}
Clearly the expectation is very badly violated, with the splitting much too
large to be described by the small current-quark masses.  What is wrong?

Chiral symmetry can be related to properties of the quark propagator, $S(p)$.
For a free quark
\begin{equation}
S_0(p)= 
\frac{ { m} - i\gamma\cdot p }{m^2 + { p}^2 }\,,
\end{equation}
with $\{\gamma_\mu$, $\mu = 1, \ldots, 4\}$ the Dirac matrices, and as a
matrix
\begin{eqnarray}
S_0(p) & \to & {\rm e}^{i \gamma_5 \theta} S_0(p) {\rm e}^{i \gamma_5 \theta}
 = \frac{- i\gamma\cdot p }{p^2 + m^2} + { \rm e}^{ 2 i \gamma_5 \theta}
\frac{{ m}}{{ p^2 + } { m}^2}\,
\end{eqnarray}
under a chiral transformation.  As anticipated, for $m=0$, $S_0(p) \to
S_0(p)$; i.e., the symmetry breaking term is proportional to the
current-quark mass and it can be measured by the ``quark condensate''
\begin{equation}
-\langle \bar q q \rangle := \int\frac{d^4 p}{(2\pi)^4}\,{\rm
tr}\left[S(p)\right] \propto \int\frac{d^4 p}{(2\pi)^4}\,\frac{m}{p^2+m^2}\,,
\end{equation}
which is the ``Cooper-pair'' density in QCD.  For a free quark the condensate
vanishes if $m=0$ but what is the effect of interactions?

Interactions dress the quark propagator so that it takes the form
\begin{equation}
\label{Sp}
S(p) := \frac{1}{i\gamma\cdot p + \Sigma(p)}
= \frac{- i\gamma\cdot p A(p^2) + B(p^2)}{p^2 A^2(p^2) + B^2(p^2) }\,,
\end{equation}
where $\Sigma(p)$ is the self energy, expressed in terms of the scalar
functions: $A$ and $B$, which are $p^2$-dependent because the interaction is
momentum-dependent.  On the valid [weak-coupling] domain they can be
calculated in perturbation theory and at one-loop order
\begin{equation}
\label{Boneloop}
B(p^2) = m \left( 1 - \frac{3\alpha_S}{4\pi} 
\ln\left[\frac{p^2}{m^2}\right] \right)\,,
\end{equation}
which is $ \propto m$.  This result persists: at every order in perturbation
theory every mass-like correction to $S(p)$ is $\propto m$ so that $m$ is
apparently the {\it only} source of chiral symmetry breaking and $\langle
\bar q q \rangle \propto m \to 0$ as $m\to 0$.  The current-quark masses are
the only explicit chiral symmetry breaking terms in QCD.

However, symmetries can be ``dynamically'' broken.  Consider a point-particle
in a rotationally invariant potential $V(\sigma,\pi) = (\sigma^2 + \pi^2 -
1)^2$, where $(\sigma,\pi)$ are the particle's coordinates.
\begin{figure}[t]

\vspace*{-6em}

\centering{\ \epsfig{figure=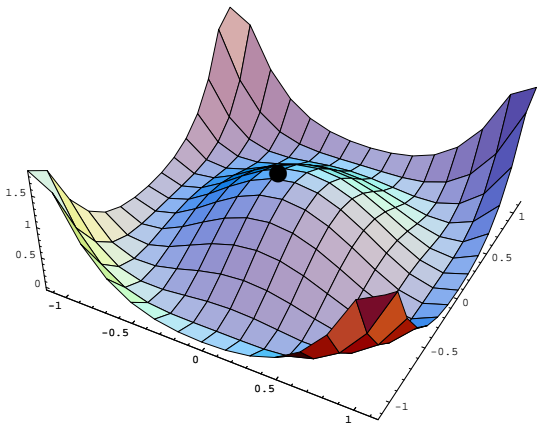,height=10.0cm}}
\vspace*{-16.5em}

{\LARGE $\bullet$}\vspace*{8em}

\caption{A rotationally invariant but unstable extremum of the Hamiltonian
obtained with the potential $V(\sigma,\pi)= (\sigma^2 + \pi^2 - 1)^2$.
\label{ball}}
\end{figure}
In the state depicted in Fig.~\ref{ball}, the particle is stationary at an
extremum of the action that is rotationally invariant but unstable.  In the
ground state of the system, the particle is stationary at any point
$(\sigma,\pi)$ in the trough of the potential, for which $\sigma^2+\pi^2=1$.
There are an uncountable infinity of such vacua, $|\theta\rangle$, which are
related one to another by rotations in the $(\sigma,\pi)$-plane.  The vacua
are degenerate but not rotationally invariant and hence, in general, $\langle
\theta | \sigma | \theta\rangle \neq 0$.  In this case the rotational
invariance of the Hamiltonian is not exhibited in any single ground state:
the symmetry is dynamically broken with interactions being responsible for
$\langle \theta | \sigma | \theta\rangle \neq 0$.
\vspace*{0.5em}

\begin{figure}[t] 
\centering{\ \epsfig{figure=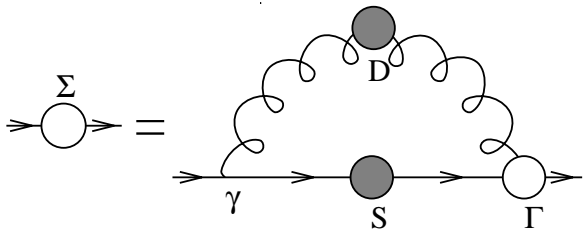,height=3.0cm} }
\caption{DSE for the dressed-quark self-energy.  The kernel of this equation
is constructed from the dressed-gluon propagator ($D$ - spring) and the
dressed-quark-gluon vertex ($\Gamma$ - open circle).  One of the vertices is
bare (labelled by $\gamma$) as required to avoid over-counting.
\label{quarkdse}}
\end{figure}

\hspace*{-\parindent}{\it \arabic{section}.\arabic{subsection}.2}~{\bf
Dynamical chiral symmetry breaking}.~The analogue in QCD is $\langle \bar q
q\rangle \neq 0$ when $m=0$.  At any finite order in perturbation theory that
is impossible.  However, using the Dyson-Schwinger equation [DSE] for the
quark self energy [the QCD ``gap equation'']:
\begin{eqnarray}
\label{qdse}
\lefteqn{i\gamma\cdot p \,A(p^2) + B(p^2) 
= Z_2\,i\gamma\cdot p \,+ Z_4\,m}\\
&& \nonumber + Z_1\, \int^\Lambda {d^4\ell\over (2\pi)^4}\,
g^2\,D_{\mu\nu}(p-\ell)\, \gamma_\mu\frac{\lambda^a}{2} \frac{1}{i\gamma\cdot
\ell A(\ell^2) + B(\ell^2)} \Gamma_\nu^a(\ell,p)\,,
\end{eqnarray}
depicted in Fig.~\ref{quarkdse}, 
it is possible to sum infinitely many contributions.\footnote{
In Eq.~(\protect\ref{qdse}), $m$ is the $\Lambda$-dependent current-quark
bare mass and $\int^\Lambda$ represents mnemonically a {\em
translationally-invariant} regularisation of the integral, with $\Lambda$ the
regularisation mass-scale.  The final stage of any calculation is to remove
the regularisation by taking the limit $\Lambda \to \infty$.  The
quark-gluon-vertex and quark wave function renormalisation constants,
$Z_1(\zeta^2,\Lambda^2)$ and $Z_2(\zeta^2,\Lambda^2)$, depend on the
renormalisation point, $\zeta$, and the regularisation mass-scale, as does
the mass renormalisation constant $Z_m(\zeta^2,\Lambda^2) :=
Z_2(\zeta^2,\Lambda^2)^{-1} Z_4(\zeta^2,\Lambda^2)$.}
That allows one to expose effects in QCD which are inaccessible in
perturbation theory.

The quark DSE is a nonlinear integral equation for $A$ and $B$ and its
nonlinearity is what makes it possible to generate nonperturbative effects.
The kernel of the equation is composed of the dressed-gluon propagator:
\begin{equation}
\label{gprop}
g^2 D_{\mu\nu}(k)  = \left(\delta_{\mu\nu} - \frac{k_\mu k_\nu}{k^2}\right)
        \frac{{\cal G}(k^2)}{k^2}\,,\; 
        {\cal G}(k^2):= \frac{g^2}{[1+\Pi(k^2)]} \,,
\end{equation}
where $\Pi(k^2)$ is the vacuum polarisation, which contains all the dynamical
information about gluon propagation, and the dressed-quark-gluon vertex:
$\Gamma^a_\mu(k,p)$.  The bare (undressed) vertex is
\begin{equation}
\Gamma_\mu^a(k,p)_{\rm bare} = \gamma_\mu\,\frac{\lambda^a}{2}\,.
\end{equation}
Once $D_{\mu\nu}$ and $\Gamma^a_\mu$ are known, Eq.~(\ref{qdse}) is
straightforward to solve by iteration.  One chooses an initial seed for the
solution functions: $_0\!A$ and $_0\!B$, and evaluates the integral on the
right-hand-side (r.h.s.).  The bare propagator values: $_0\!A=1$ and $_0\!B
=m$ are often adequate.  This first iteration yields new functions: $_1\!A$
and $_1\!B$, which are reintroduced on the r.h.s. to yield $_2\!A$ and
$_2\!B$, etc.  The procedure is repeated until $_n\!A = \,_{n+1}\!A$ and
$_n\!B = \,_{n+1}\!B$ to the desired accuracy.

\begin{figure}[t] 
\vspace*{-1.0em}

\centering{\ \epsfig{figure=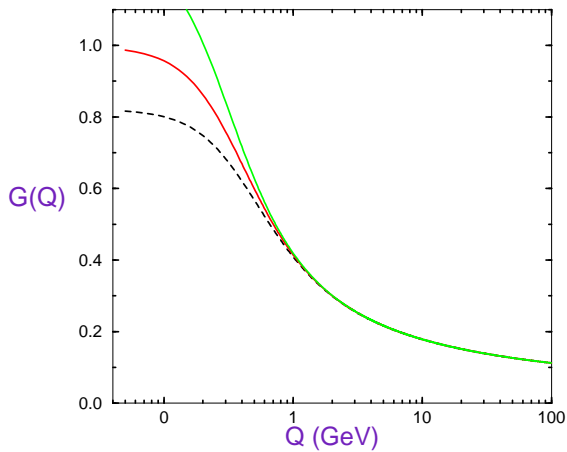,height=7.0cm}}
\caption{\label{GQ} Illustrative forms of ${\cal G}(Q)$ that agree with the
perturbative result at $Q>1\,$GeV.  Three possibilities are canvassed: ${\cal
G}(Q=0)<1$; ${\cal G}(Q=0)= 1$; and ${\cal G}(Q=0)> 1$.}
\end{figure}

It is now simple to illustrate DCSB.  Using the bare vertex and
${\cal G}(Q)$ depicted in Fig.~\ref{GQ}, I solved
the quark DSE in the chiral limit.  If ${\cal G}(Q=0)<1$ then $B(p^2)\equiv
0$ is the only solution.  However, when ${\cal G}(Q=0)\geq 1$ the equation
admits an energetically favoured $B(p^2)\not\equiv 0$ solution; i.e., if the
coupling is large enough then even in the absence of a current-quark mass,
contrary to Eq.~(\ref{Boneloop}), the quark acquires a mass {\it dynamically}
and hence
\begin{equation}
\langle \bar q q \rangle  \propto \int\frac{d^4 p}{(2\pi)^4}\,
\frac{B(p^2)}{p^2\,A(p^2)^2+ B(p^2)^2} \neq 0\;\; {\rm for}\; m=0\,.
\end{equation}

This identifies a mechanism for DCSB in quantum field theory.  The nonzero
condensate provides a new, dynamically generated mass-scale and if its
magnitude is large enough$\,$\footnote{$-\langle \bar q q \rangle^{1/3}$ need
only be one order-of-magnitude larger than $m_u\sim m_d$.}  it can explain
the mass splitting between parity partners, and many other surprising
phenomena in QCD.  The simple model illustrates that DCSB is linked to the
long-range behaviour of the fermion-fermion interaction and the same is true
of confinement.  The question is then: How does $D_{\mu\nu}(k)$ behave in
QCD?
\section{Dyson-Schwinger Equations}
In the last section I introduced the DSE for the quark self energy.  It is
one of an infinite tower of coupled integral equations, with the equation for
a particular $n$-point function involving at least one $m>n$-point function;
e.g., the quark DSE involved the dressed-gluon propagator, a 2-point
function, and the dressed-quark-gluon vertex, a 3-point function.  The
collection of DSEs provide a Poincar\'e invariant, continuum approach to
solving quantum field theories.  However, as an infinite collection of
coupled equations, a tractable problem is only obtained if one truncates the
system.  Historically this has provided an impediment to the application of
DSEs: {\it a priori} it can be difficult to judge whether a particular
truncation scheme will yield qualitatively or quantitatively reliable results
for the quantity sought.  As we saw, the DSEs are integral equations and
hence the analysis of observables is a numerical problem.  Therefore a
critical evaluation of truncation schemes often requires access to high-speed
computers.\footnote{The human and computational resources required are still
modest compared with those consumed in contemporary numerical simulations of
lattice-QCD.}  With such tools now commonplace, this evaluation can be
pursued fruitfully.

The development of efficacious truncation schemes is not a purely numerical
task, and neither is it always obviously systematic.  For some, this last
point diminishes the appeal of the approach.  However, with growing community
involvement and interest, the qualitatively robust results and intuitive
understanding that the DSEs can provide is becoming clear.  Indeed, those
familiar with the application of DSEs in the late-70s and early-80s might be
surprised with the progress that has been made.  It is now
clear$\,$\cite{bender96,QC96} that truncations which preserve the global
symmetries of a theory; e.g., chiral symmetry in QCD, are relatively easy to
define and implement and, while it is more difficult to preserve local gauge
symmetries, much progress has been made with Abelian
theories$\,$\cite{ayse97} and more is being learnt about non-Abelian ones.

The simplest truncation scheme for the DSEs is the weak-coupling expansion,
which shows that they {\it contain} perturbation theory; i.e, for any given
theory the weak-coupling expansion generates all the diagrams of perturbation
theory.  However, as with DCSB, the most important feature of the DSEs is the
antithesis of this weak-coupling expansion: the DSEs are intrinsically
nonperturbative and their solution contains information that is {\it not}
present in perturbation theory.  They are ideal for studying the phenomena I
identified as the core of these lectures and in this application they provide
a means of elucidating identifiable signatures of the quark-gluon
substructure of hadrons.
\subsection{Gluon Propagator}\label{subsec:B1}
\begin{figure}[t] 
\hspace*{30mm}
Charge Screening  \hspace*{2.0cm} 
        Charge AntiScreening \\
\hspace*{42mm} $ \searrow$ \hspace*{43mm} $ \swarrow$ 

\centering{\ \epsfig{figure=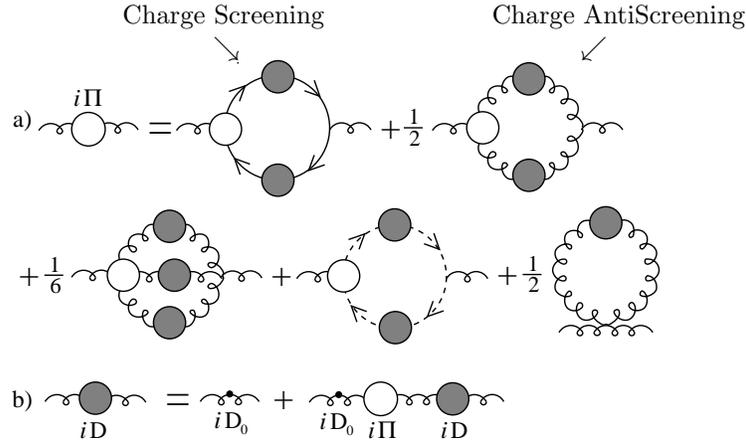,height=5.0cm} }
\caption{DSE for the gluon vacuum polarisation and propagator: solid line -
quark; spring - gluon; dotted-line - ghost.  The open circles are irreducible
vertices.  As indicated, the quark loop acts to screen the charge, as in QED,
while the gluon loop opposes this, anti-screening the charge and enhancing
the interaction.
\label{gluondse}}
\end{figure}

In Landau gauge the dressed-gluon propagator has the form in
Eq.~(\ref{gprop}) and satisfies the DSE [a nonlinear integral equation]
depicted in Fig.~\ref{gluondse}.
A weak coupling expansion of the equation reproduces perturbation theory and
shows directly that in the one-loop expression for the running coupling
constant, Eq.~(\ref{alphaQCD}), the ``$11 N_c$'' comes from the
charge-antiscreening gluon loop and the ``$2 N_f$'' from the charge-screening
fermion loop.  This illustrates how the non-Abelian structure of QCD is
responsible for asymptotic freedom and suggests that confinement is related
to the importance of gluon self-interactions.

Studies of the gluon DSE have been reported by many authors$\,$\cite{rw94}
with the conclusion that if the ghost-loop is unimportant, then the
charge-antiscreening 3-gluon vertex dominates and, relative to the free gauge
boson propagator, the dressed gluon propagator is significantly enhanced in
the vicinity of $k^2=0$.  The enhancement persists to $k^2 \sim
1$-$2\,$GeV$^2$, where a perturbative analysis becomes quantitatively
reliable.  In the neighbourhood of $k^2=0$ the enhancement can be
represented$\,$\cite{bp89} as a regularisation of $1/k^4$ as a distribution.
A dressed-gluon propagator of this type generates confinement and DCSB {\it
without} fine-tuning, as I will elucidate.
\subsection{Quark Propagator}\label{subsec:B2}
\begin{figure}[t]
\vspace*{-1.0em}

\centering{\ 
\epsfig{figure=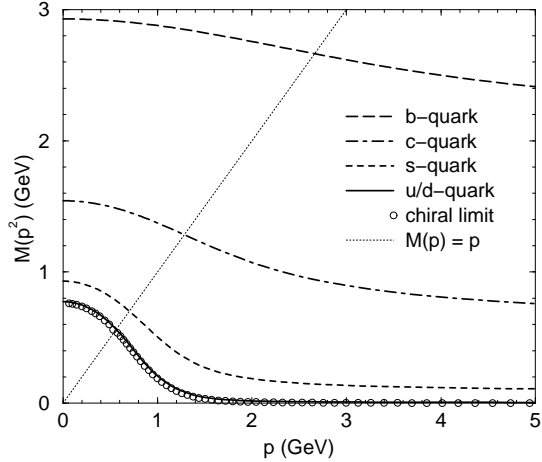,height=7.0cm}}
\caption{Dressed-quark mass-function obtained in solving the quark DSE.
\label{plotMpp}}
\end{figure}

In a covariant gauge the dressed-quark propagator can be written in a number
of equivalent forms: those in Eq.~(\ref{Sp}) and also as 
\begin{equation}
\label{Spsv}
S(p) =\, - i\gamma\cdot p \,\sigma_V(p^2) + \sigma_S(p^2) 
 = \frac{Z(p^2)}{i\gamma\cdot p + M(p^2)}\,.
\end{equation}
As depicted in Fig.~\ref{plotMpp}, 
solving the quark DSE, Eq.~(\ref{qdse}), using a dressed-gluon propagator of
the type described above, one obtains a quark mass-function, $M(p^2)$, that
mirrors the enhancement of the dressed-gluon propagator.  The results in the
figure were obtained$\,$\cite{mr97} with current-quark masses corresponding
to
\begin{equation}
\label{monegev}
\begin{array}{lllllll}
m_{u/d}^{1\,{\rm GeV}} &\;&
m_s^{1\,{\rm GeV}}     & \;&
m_c^{1\,{\rm GeV}}     &\; &
m_b^{1\,{\rm GeV}}     \\
 6.6\, {\rm MeV} &\; &
 140\,{\rm MeV}  &\;& 
 1.0\,{\rm GeV}  &\; &
 3.4\,{\rm GeV}\,.
\end{array}
\end{equation}

The quark DSE was also solved in the chiral limit, which in QCD is obtained
by setting the Lagrangian current-quark bare mass to zero.\cite{mr97} One
observes immediately that the mass-function is nonzero even in this case.
That {\it is} DCSB: a momentum-dependent quark mass, generated dynamically,
in the absence of any term in the action that breaks chiral symmetry
explicitly.  This entails a nonzero value for the quark condensate in the
chiral limit.  That $M(p^2)\neq 0$ in the chiral limit is independent of the
details of the infrared enhancement in the dressed-gluon propagator.

Figure~\ref{plotMpp} illustrates that for light quarks ($u$, $d$ and $s$)
there are two distinct domains: perturbative and nonperturbative.  In the
perturbative domain the magnitude of $M(p^2)$ is governed by the the
current-quark mass. For $p^2< 1\,$GeV$^2$ the mass-function rises sharply.
This is the nonperturbative domain where the magnitude of $M(p^2)$ is
determined by the DCSB mechanism; i.e., the enhancement in the dressed-gluon
propagator.  This emphasises that DCSB is more than just a nonzero value of
the quark condensate in the chiral limit!  

The solution of $p^2=M^2(p^2)$ defines a Euclidean constituent-quark mass,
$M^E$.  For a given quark flavour, the ratio ${\cal L}_f:=M^E_f/m_f^\zeta$ is
a single, quantitative measure of the importance of the DCSB mechanism in
modifying the quark's propagation characteristics.  As illustrated in
Eq.~(\ref{Mmratio}),
\begin{equation}
\label{Mmratio}
\begin{array}{l|c|c|c|c|c}
\mbox{\sf flavour} 
        &   u/d  &   s   &  c  &  b  &  t \\\hline
 \frac{M^E}{m^{\zeta\sim 20\,{\rm GeV}}}
       &  150   &    10      &  2.3 &  1.4 & \to 1
\end{array}\,,
\end{equation}
this ratio provides for a natural classification of quarks as either light or
heavy.  For light-quarks ${\cal L}_f$ is characteristically $10$-$100$ while
for heavy-quarks it is only $1$-$2$.  The values of ${\cal L}_f$ signal the
existence of a characteristic DCSB mass-scale: $M_\chi$. At $p^2>0$ the
propagation characteristics of a flavour with $m_f^\zeta< M_\chi$ are altered
significantly by the DCSB mechanism, while for flavours with $m_f^\zeta\gg
M_\chi$ it is irrelevant, and explicit chiral symmetry breaking dominates.
It is apparent from Eq.~(\ref{Mmratio}) that $M_\chi \sim 0.2\,$GeV$\,\sim
\Lambda_{\rm QCD}$.  This forms a basis for many simplifications in the study
of heavy-meson observables.\cite{misha}

\subsection{Confinement}\label{subsec:conf}
Confinement is the absence of quark and gluon production thresholds in
colour-singlet-to-singlet ${\cal S}$-matrix amplitudes.  That is ensured if
the dressed-quark and -gluon propagators do not have a Lehmann
representation.  

To illustrate a Lehmann representation, consider the 2-point free-scalar
propagator: $ \Delta(k^2)= 1/[k^2+m^2]$.  One can write
\begin{equation}
\Delta(z)= \int_0^\infty\,d\sigma\, \frac{\rho(\sigma)}{z + \sigma}\,,
\end{equation}
where in this case the spectral density is 
\begin{equation}
\rho(x):= \frac{1}{2 \pi i} \lim_{\epsilon \to 0} 
\left[\Delta(-x - i\epsilon ) - \Delta(-x + i\epsilon )\right]
= \delta(m^2-x)\,,
\end{equation}
which is non-negative.  That is a Lehmann representation: each scalar
function necessary to specify the $n$-point function completely has a
spectral decomposition with non-negative spectral densities.  Only those
functions whose poles or branch points lie at timelike, real-$k^2$ have a
Lehmann representation.

The existence of a Lehmann representation for a dressed-particle propagator
is necessary if the construction of asymptotic ``in'' and ``out'' states for
the associated quanta is to proceed; i.e., it is necessary if these quanta
are to propagate to a ``detector''.  In its absence there are no asymptotic
states with the quantum numbers of the field whose propagation
characteristics are described by the propagator.  Structurally, the
nonexistence of a Lehmann representation for the dressed-propagators of
elementary fields ensures the absence of pinch singularities in loops and
hence the absence of quark and gluon production thresholds.

The mechanism can be generalised and applied to coloured bound states, such
as colour-antitriplet quark-quark composites (diquarks).  A
study$\,$\cite{bender96} of the quark-quark scattering matrix shows that it
does not have a spectral decomposition with non-negative spectral densities
and hence there are no diquark bound states.  The same argument that
demonstrates the absence of diquarks in the spectrum of $SU(N_c=3)$ also
proves$\,$\cite{QC96} that in $SU(N_c=2)$ the ``baryons'', which are
necessarily diquarks in this theory, are degenerate with the mesons.

Dressed-gluon propagators with the infrared enhancement described above do
not have a Lehmann representation and using forms like this in the kernel of
the quark DSE yields a dressed-quark propagator that also does not have a
Lehmann representation.  In this sense confinement {\it breeds} confinement,
without fine-tuning.
\subsection{Hadrons: Bound States}\label{subsec:B3}
The properties of hadrons can be understood in terms of their substructure by
studying covariant bound state equations: the Bethe-Salpeter equation [BSE]
for mesons and the covariant Fadde'ev equation for baryons.  The mesons have
been studied most extensively and their internal structure is described by a
Bethe-Salpeter amplitude obtained as a solution of
\begin{eqnarray}
\label{genbse}
\left[\Gamma_H(k;P)\right]_{tu} &= & 
\int^\Lambda \frac{d^4q}{(2\pi)^4}\,
[\chi_H(q;P)]_{sr} \,K^{rs}_{tu}(q,k;P)\,,
\end{eqnarray}
where $\chi_H(q;P) := {\cal S}(q_+) \Gamma_H(q;P) {\cal S}(q_-)$; ${\cal
S}(q) = {\rm diag}(S_u(q),S_d(q),S_s(q), \ldots)$; $q_+=q + \eta_P\, P$,
$q_-=q - (1-\eta_P)\, P$, with $P$ the total momentum of the bound state; and
$r$,\ldots,$u$ represent colour-, Dirac- and flavour-matrix indices.  The
amplitude for a pseudoscalar bound state has the form
\begin{eqnarray}
\label{genpibsa}
\Gamma_H(k;P) & = &  T^H \gamma_5 \left[ i E_H(k;P) + 
\gamma\cdot P F_H(k;P) \rule{0mm}{5mm}\right. \\
\nonumber & & 
\left. \rule{0mm}{5mm}+ \gamma\cdot k \,k \cdot P\, G_H(k;P) 
+ \sigma_{\mu\nu}\,k_\mu P_\nu \,H_H(k;P) 
\right]\,,
\end{eqnarray}
where $T^H$ is a flavour matrix that determines the channel under
consideration; e.g., $T^{K^+}:= (1/2)\left(\lambda^4 + i \lambda^5\right)$,
with $\{\lambda^j,j=1\ldots 8\}$ the Gell-Mann matrices.

In Eq.~(\ref{genbse}), $K$ is the renormalised, fully-amputated,
quark-antiquark scattering kernel and important in the successful application
of DSEs is that it has a systematic skeleton expansion in terms of the
elementary, dressed-particle Schwinger functions; e.g., the dressed-quark and
-gluon propagators.
\begin{figure}[t]
  \centering{\ \hspace*{-2.5cm}\epsfig{figure=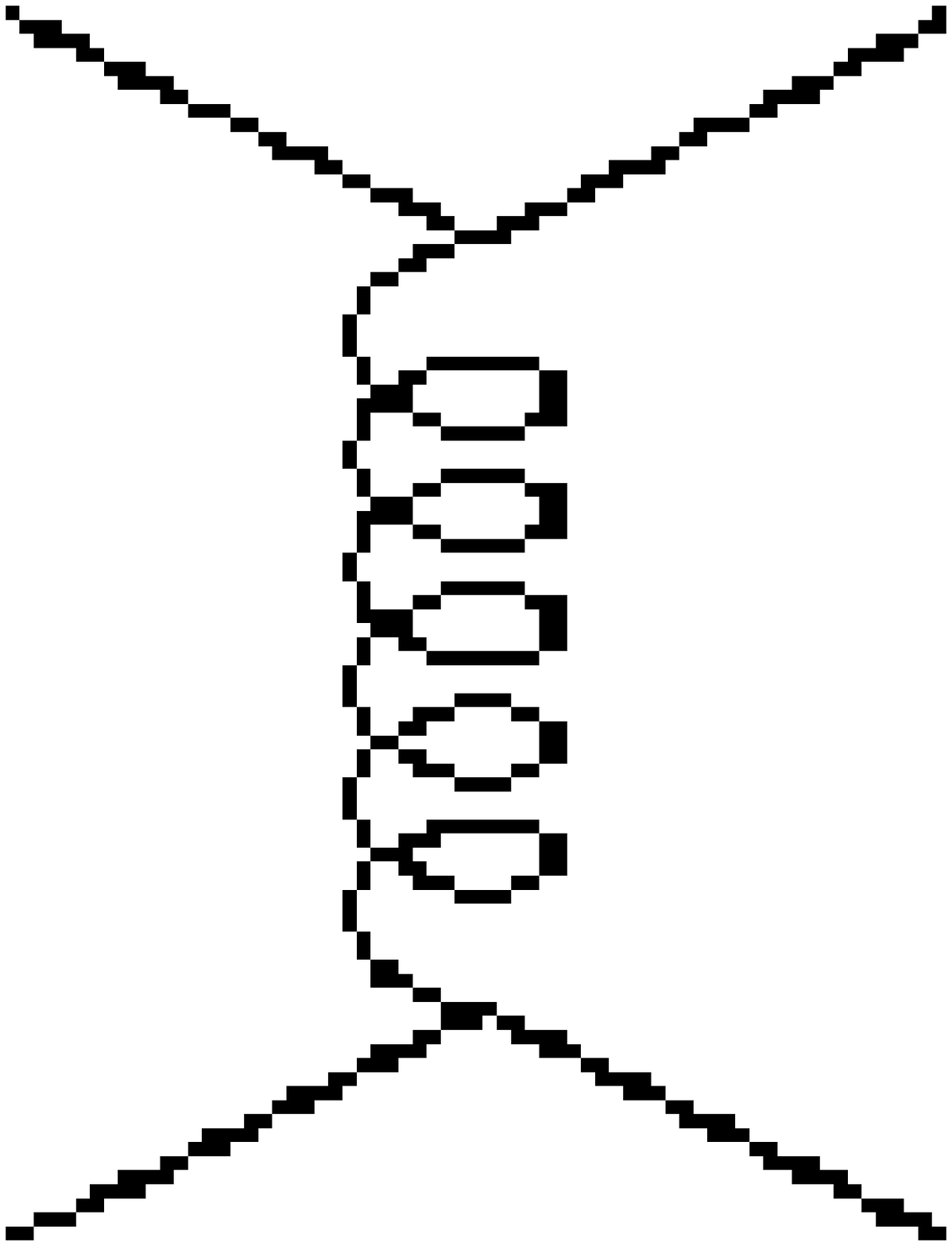,height=2.0cm} 
        \vspace*{-15mm} 

        \hspace*{15mm} $\longleftarrow$ { (1) -- Ladder}\vspace*{4mm} 

        \hspace*{55mm} { (2) -- Beyond Ladder}

        \hspace*{16mm}$\swarrow$

        \hspace*{0.5cm}\epsfig{figure=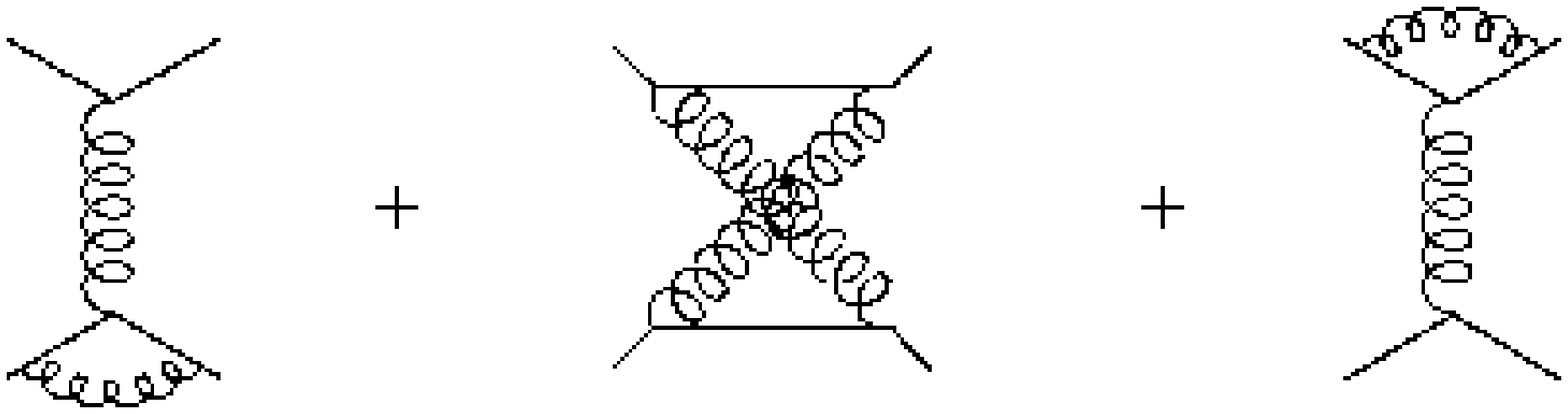,height=2.0cm} }
\caption{First two orders in a systematic expansion of the quark-antiquark
scattering kernel.  In this expansion, the propagators are dressed but the
vertices are bare.
\label{skeleton}}
\end{figure}
The expansion introduced in Ref.~[\ref{bender96R}] provides a means of
constructing a kernel that, order-by-order in the number of vertices, ensures
the preservation of vector and axial-vector Ward-Takahashi identities; i.e.,
current conservation.

In any study of meson properties, one chooses a truncation for $K$.  The BSE
is then fully specified and straightforward to solve, yielding the bound
state mass and amplitude.  The ``ladder'' truncation of $K$ combined with the
``rainbow'' truncation of the quark DSE [$\Gamma_\mu \to \gamma_\mu$ in
Eq.~(\ref{qdse})] is the simplest and most often used.  The expansion of
Fig.~\ref{skeleton} provides the explanation$\,$\cite{bender96} for why this
Ward-Takahashi identity preserving truncation is accurate for
flavour-nonsinglet pseudoscalar and vector mesons: there are cancellations
between the higher-order diagrams.  It also shows why it provides a poor
approximation in the study of scalar mesons, where the higher-order terms do
not cancel, and for flavour-singlet mesons, where it omits timelike gluon
exchange diagrams.

\section{A Bound State Mass Formula}\label{sec:MF}
The dressed-axial-vector vertex satisfies a DSE whose kernel is $K$, and
because of the systematic expansion described in Sec.~\ref{subsec:B3} it
follows$\,$\cite{mr97} that the axial-vector Ward-Takahashi identity
[AV-WTI]:
\begin{eqnarray}
\label{avwti}
\lefteqn{-i P_\mu \Gamma_{5\mu}^H(k;P) = }\\ && \nonumber {\cal
S}^{-1}(k_+)\gamma_5\frac{T^H}{2} + \gamma_5\frac{T^H}{2} {\cal S}^{-1}(k_-)
- M_{(\zeta)}\,\Gamma_5^H(k;P) - \Gamma_5^H(k;P)\,M_{(\zeta)} \,,
\end{eqnarray}
[$M_{(\zeta)}= {\rm diag}(m_u^\zeta,m_d^\zeta,m_s^\zeta,\ldots)$ is the
current-quark mass matrix] is satisfied in any thoughtful truncation of the
DSEs.  That entails many important results.

{\boldmath $f_H$}:~~The axial-vector vertex has a pole
at $P^2=-m_H^2$ whose residue is $f_H$, the leptonic decay constant:
\begin{eqnarray}
\label{caint}
f_H P_\mu = 
Z_2\int^\Lambda \frac{d^4q}{(2\pi)^4}\,\case{1}{2}
{\rm tr}\left[\left(T^H\right)^{\rm t} \gamma_5 \gamma_\mu 
{\cal S}(q_+) \Gamma_H(q;P) {\cal S}(q_-)\right]\,,
\end{eqnarray} 
with the trace over colour, Dirac and flavour indices.  This expression is
exact: the dependence of $Z_2$ on the renormalisation point, regularisation
mass-scale and gauge parameter is just that necessary to ensure that the
l.h.s. is independent of all these things.

{\bf Goldstone's Theorem}:~~In the chiral limit
\begin{equation}
\label{bwti}
\begin{array}{ll}
f_H E_H(k;0)  =   B_0(k^2)\,,\; & 
F_R(k;0) +  2 \, f_H F_H(k;0)   =  A_0(k^2)\,, \\
G_R(k;0) +  2 \,f_H G_H(k;0)     =  2 A_0^\prime(k^2)\,,\; &
H_R(k;0) +  2 \,f_H H_H(k;0)     =  0\,,
\end{array}
\end{equation}
where $A_0(k^2)$ and $B_0(k^2)$ are the solutions of Eq.~(\ref{qdse}) in the
chiral limit, and $F_R$, $G_R$ and $H_R$ are calculable functions in
$\Gamma^H_{5\mu}$.  This shows that when chiral symmetry is dynamically
broken: 1) the flavour-nonsinglet, pseudoscalar BSE has a massless solution;
2) the Bethe-Salpeter amplitude for the massless bound state has a term
proportional to $\gamma_5$ alone, with the momentum-dependence of $E_H(k;0)$
completely determined by that of $B_0(k^2)$, in addition to terms
proportional to other pseudoscalar Dirac structures that are nonzero in
general; and 3) the axial-vector vertex, $\Gamma_{5 \mu}^H(k;P)$, is
dominated by the pseudoscalar bound state pole for $P^2\simeq 0$.  The
converse is also true.  Hence, in the chiral limit, the pion is a massless
composite of a quark and an antiquark, each of which has an effective mass
$M^E \sim 450\,$MeV.  With an infrared enhanced dressed-gluon propagator of
the type described in Sec.~\ref{subsec:B1}, this occurs without fine-tuning.

{\bf In-meson Condensate}:~~The pseudoscalar vertex also has a pole at
$P^2=-m_H^2$ whose residue is
\begin{equation}
\label{rH}
i r_H = Z_4\int^\Lambda \frac{d^4q}{(2\pi)^4}\,\case{1}{2} {\rm
tr}\left[\left(T^H\right)^{\rm t} \gamma_5 {\cal S}(q_+) \Gamma_H(q;P) {\cal
S}(q_-)\right]:= - \,\frac{\langle \bar q q \rangle^H_\zeta}{f_H}\,.
\end{equation}
The renormalisation constant $Z_4$ on the r.h.s. depends on the gauge
parameter, the regularisation mass-scale and the renormalisation point.  This
dependence is exactly that required to ensure $r_H$ is finite in the limit
$\Lambda\to \infty$ and gauge-parameter independent.  $\langle \bar q q
\rangle^H_\zeta$ is the in-meson quark condensate.

{\bf Mass Formula}:~~There is an identity between the residues of the
pseudoscalar meson pole in the axial-vector and pseudoscalar vertices that is
satisfied independent of the magnitude of the current-quark mass:
\begin{equation}
\label{gmora}
f_H^2\,m_H^2 = \,- \,{\cal M}_H\,\langle \bar q q \rangle^H_\zeta \, ,\;\;
{\cal M}_H := {\rm tr}_{\rm flavour}
\left[M_{(\zeta)}\,\left\{T^H,\left(T^H\right)^{\rm t}\right\}\right]\,,
\end{equation}
e.g., for the $\pi$: ${\cal M}_H = m_u^\zeta + m_d^\zeta$.  This is a mass
formula for flavour-octet pseudoscalar mesons and the r.h.s. does not involve
a difference of massive quark propagators: a phenomenological assumption
often employed.  The renormalisation point dependence of $\langle \bar q q
\rangle^H_\zeta$ is exactly such that the r.h.s. of Eq.~(\ref{gmora}) is
renormalisation point {\it independent}.
\subsection{A Corollary}
For small current-quark masses, using Eqs.~(\ref{genpibsa}) and
(\ref{bwti}), Eq.~({\ref{rH}) yields
\begin{equation}
\label{cbqbq}
\begin{array}{lcr}
\displaystyle
r_H^0  =  -\,\frac{1}{f_H^0}\, \langle \bar q q \rangle_\zeta^0 \,
, & & 
\displaystyle
\,-\,\langle \bar q q \rangle_\zeta^0 :=  
Z_4(\zeta^2,\Lambda^2)\, N_c \int^\Lambda_q\,{\rm tr}_{\rm Dirac}
        \left[ S_{\hat m =0}(q) \right]\,,
\end{array}
\end{equation}
where the superscript ``$0$'' denotes that the quantity is evaluated in the
chiral limit and $ \langle \bar q q \rangle_\zeta^0 $, as defined here, is the
chiral limit {\it vacuum quark condensate}; i.e., the in-meson condensate
becomes the vacuum quark condensate in the chiral limit.  One obtains
immediately from Eqs.~(\ref{gmora}) and (\ref{cbqbq})
\begin{eqnarray}
\label{gmorepi}
f_{\pi}^2 m_{\pi}^2 & = &-\,\left[m_u^\zeta + m_d^\zeta\right]
       \langle \bar q q \rangle_\zeta^0 + {\rm O}\left(\hat m_q^2\right)\,,\\
\label{gmoreKp}
f_{K^+}^2 m_{K^+}^2 & = &-\,\left[m_u^\zeta + m_s^\zeta\right]
       \langle \bar q q \rangle_\zeta^0 + {\rm O}\left(\hat m_q^2\right)\,,
\end{eqnarray}
which exemplify what is commonly known as the Gell-Mann--Oakes--Renner
relation.  [$\hat m_q$ is the renormalisation-point-independent current-quark
mass.]  Since Eq.~(\ref{gmora}) is valid {\it independent} of the
current-quark mass, it is also applicable to heavy pseudoscalar mesons.  A
corollary valid in this sector is described in Sec.~\ref{sec:hq}.
\section{An Exemplar}\label{sec:model}
I have already made use of a model$\,$\cite{mr97} to illustrate some of the
robust results of DSE studies and I will now describe and explore that model
in more detail.  For the kernel of the quark DSE it employs the analogue of
the lowest-order BSE kernel in Fig.~\ref{skeleton}:
\begin{eqnarray}
\label{ouransatz}
\lefteqn{Z_1\, \int^\Lambda \frac{d^4 q}{(2\pi)^4} \,
g^2 D_{\mu\nu}(p-q) \frac{\lambda^a}{2}\gamma_\mu S(q)
\Gamma^a_\nu(q,p)}\\
& & \nonumber \to
\int^\Lambda \frac{d^4 q}{(2\pi)^4}\,
{\cal G}((p-q)^2)\, D_{\mu\nu}^{\rm free}(p-q)
 \frac{\lambda^a}{2}\gamma_\mu S(q)
\frac{\lambda^a}{2}\gamma_\nu \,.
\end{eqnarray}
This is the ``rainbow'' approximation, in which the specification of the
model is complete once a form is chosen for the ``effective coupling'' ${\cal
G}(k^2)$.

The requirement of consistency with the AV-WTI motivates the {\it
Ansatz}$\,$:
\begin{equation}
\label{abapprox}
{\cal G}(Q^2):= 4\pi\,\alpha(Q^2)\,,
\end{equation}
so that the form of ${\cal G}(Q^2)$ at large-$Q^2$ is fixed by that of the
running coupling constant.  This {\it Ansatz} is often described as the
``Abelian approximation'' because the l.h.s. equals the r.h.s. in QED.  In
QCD, equality between the two sides of Eq.~(\ref{abapprox}) cannot be
obtained easily by a selective resummation of diagrams.  As reviewed in
Ref.~[\ref{rw94R}], Eqs.~(5.1) to (5.8), it can only be achieved by enforcing
equality between the renormalisation constants for the ghost-gluon vertex and
ghost wave function: $\tilde Z_1=\tilde Z_3$.

The explicit form of the {\it Ansatz} employed in Ref.~[\ref{mr97R}] is
depicted in Fig.~\ref{Gmodel}
\begin{figure}[t]
\centering{\
\epsfig{figure=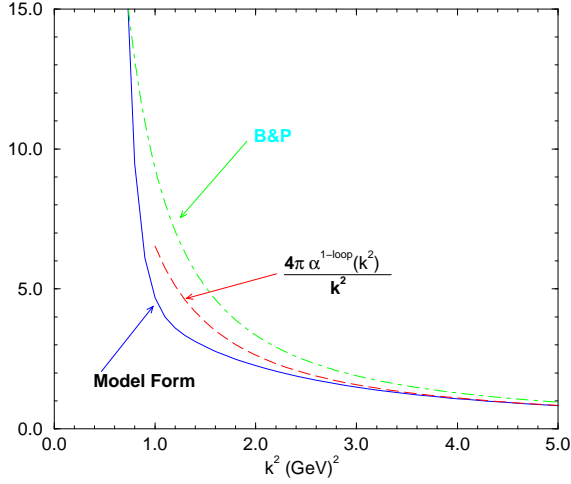,height=6.5cm}}\vspace*{-1.0\baselineskip}
\caption{{\it Ansatz}$\,$ for ${\cal G}(k^2)/k^2$ employed in
Ref.~[\protect\ref{mr97R}].  ``B\&P'' labels a
solution$\,$\protect\cite{bp89} of the gluon DSE, which is presented for
comparison, as is the one-loop running coupling in QCD.
\label{Gmodel}}
\end{figure}
and its qualitative features are easily understood.  In the infrared it has
an integrable singularity$\,$\cite{mn83} $ \propto \delta^4(k)$ and a
finite-width approximation to $\delta^4(k)$, normalised such that it has the
same $\int d^4k$ as the first term, and in the ultraviolet it is dominated by
$\alpha(k^2)/k^2$.  However, it does not have a singularity on the real-$k^2$
axis, which ensures gluon confinement through the absence of a Lehmann
representation, Sec.~\ref{subsec:conf}.

The model has ostensibly three parameters: $D$, a mass-scale, and $\omega$
and $m_t$, two range parameters.  However, in the numerical studies the
values $\omega=0.3\,$GeV$\,[=1/(.66\,{\rm fm})]$ and
$m_t=0.5\,$GeV$\,[=1/(.39\,{\rm fm})]$ were fixed, and only $D$ and the
renormalised $u/d$- and $s$-current-quark masses varied in order to satisfy
the goal of a good description of low-energy $\pi$- and $K$-meson properties.
This was achieved with
\begin{equation}
\label{params}
\begin{array}{ccc}
D= 0.781\,{\rm GeV}^2\,,\; &
m_{u/d}^\zeta = 3.74\,{\rm MeV}\,,\; &
m_s^\zeta = 82.5\,{\rm MeV}
\end{array}\,,
\end{equation}
at $\zeta\approx 20\,$GeV, which is large enough to be in the perturbative
domain.  

As remarked in Sec.~\ref{subsec:AC}, the chiral limit is unambiguously
defined by $\hat m = 0$ so there is no perturbative contribution to the
scalar piece of the quark self energy, $B(p^2)$; and in fact there is no
scalar, mass-like divergence in the perturbative calculation of the self
energy.

Figure~\ref{spplot} depicts the dressed-quark mass function, $M(p^2)$,
obtained by solving the quark DSE using the parameters in Eq.~(\ref{params}),
and in the chiral limit.
\begin{figure}[t]
\vspace*{-1.0em}

\centering{\
\epsfig{figure=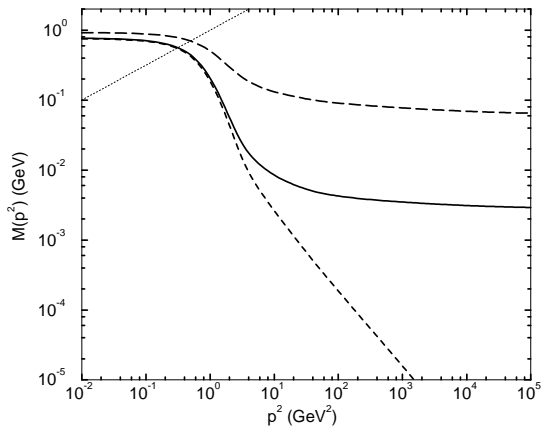,height=6.5cm}}\vspace*{-1.5\baselineskip}
\caption{Dressed-quark mass function obtained in solving the quark DSE using
the parameters in Eq.~(\protect\ref{params}): $u/d$-quark (solid line);
$s$-quark (long-dashed line); chiral limit (dashed line).  The
renormalisation point is $\zeta\approx 20\,$GeV.  The intersection of the
line $M^2(p^2)=p^2$ (dotted line) with each curve defines $M^E$, the
Euclidean constituent-quark mass.
\label{spplot}}
\end{figure}
It complements Fig.~\ref{plotMpp} by highlighting the qualitative difference
between the behaviour of $M(p^2)$ in the chiral limit and in the presence of
explicit chiral symmetry breaking.
In the latter case
\begin{equation}
\label{masanom}
 M(p^2) \stackrel{{\rm large}-p^2}{=} \frac{ \hat m}
{\left(\mbox{\footnotesize$\frac{1}{2}$}\ln\left[\frac{p^2}{\Lambda_{\rm QCD}^2}
\right]\right)^{\gamma_m}} \left\{ 1 + {\rm two~loop}\right\}\,, \;
\gamma_m=\frac{12}{33- 2 N_f}\,.
\end{equation}
However, in the chiral limit the ultraviolet behaviour is
\begin{equation}
\label{Mchiral}
M(p^2) \stackrel{{\rm large}-p^2}{=}\,
\frac{2\pi^2\gamma_m}{3}\,\frac{\left(-\,\langle \bar q q \rangle^0\right)}
           {p^2
        \left(\case{1}{2}\ln\left[\frac{p^2}{\Lambda_{\rm QCD}^2}\right]
        \right)^{1-\gamma_m}}\,,
\end{equation}
where $\langle \bar q q \rangle^0$ is the renormalisation-point-independent
vacuum quark condensate.  Analysing the chiral limit solution yields
\begin{equation}
\label{qbqM0}
-\,\langle \bar q q \rangle^0 = (0.227\,{\rm GeV})^3\,,
\end{equation}
which is a reliable means of determining $\langle \bar q q \rangle^0$ because
corrections to Eq.~(\ref{Mchiral}) are suppressed by powers of $\Lambda_{\rm
QCD}^2/\zeta^2$.

Equation~(\ref{cbqbq}) defines the renormalisation-point-dependent vacuum
quark condensate
\begin{equation}
\label{qbq19}
\left.-\,\langle \bar q q \rangle_\zeta^0 \right|^{\zeta=19\,{\rm GeV}}:=  
\left(\lim_{\Lambda\to\infty}
\left. Z_4(\zeta,\Lambda)\, N_c \int^\Lambda_q\,{\rm tr}_{\rm Dirac}
        \left[ S_{\hat m =0}(q) \right]\right)\right|^{\zeta=19\,{\rm GeV}}
\end{equation}
and the calculated$\,$\cite{mr97} value is $(0.275\,{\rm GeV})^3$.  In the
model it is straightforward to establish explicitly that $ m^\zeta\,\langle
\bar q q \rangle_\zeta^0 =\,$constant, independent of $\zeta$, and hence
\begin{equation}
\label{rgiprod}
m^\zeta\,\langle \bar q q \rangle_\zeta^0 :=  \hat m \,\langle \bar q q
\rangle^0\,,
\end{equation}
which defines the renormalisation-point-independent current-quark masses
unambiguously.  From this and Eqs.~(\ref{params}), (\ref{qbqM0}) and
(\ref{qbq19}) one obtains 
\begin{equation}
\label{rgimass}
\hat m_{u/d} = 6.60 \, {\rm MeV} \,,\;
\hat m_s = 147\, {\rm MeV}\,.
\end{equation}
Using the one-loop evolution in Eq.~(\ref{masanom}) these values yield
$m_{u/d}^\zeta= 3.2\,$MeV and $m_s^\zeta= 72\,$MeV, which are within $\sim 10$\%
of the actual values in Eq.~(\ref{params}).  This indicates that higher-loop
corrections to the one-loop formulae provide contributions of $<10$\% at $p^2
= \zeta^2$.  The magnitude of the higher-loop contributions decreases with
increasing $p^2$.

The renormalisation-point-invariant product in Eq.~(\ref{rgiprod}) also
yields 
\begin{equation}
\label{qbq1}
\left.-\,\langle \bar q q
\rangle_\zeta^0\right|_{\zeta=1\,{\rm GeV}}
:= \left(\ln\left[1/\Lambda_{\rm QCD}\right]\right)^{\gamma_m}
\, \langle \bar q q \rangle^0
= (0.241\,{\rm GeV})^3\,,
\end{equation}
which can be compared directly with the value of the quark condensate
employed in contemporary phenomenological studies:$\,$\cite{derek} $
(0.236\pm 0.008\,{\rm GeV})^3$.  It is now straightforward to determine the
accuracy of Eqs.~(\ref{gmorepi}) and (\ref{gmoreKp}).  Using experimental
values on the left-hand-side, one finds:
\begin{eqnarray}
\label{picf}
(92.4 \times 138.5)^2 = (113\,{\rm MeV})^4 & \;{\rm cf.} \;& 
(111)^4 = 2\times 5.5 \times 241^3\,, \\
\label{Kcf}
(113 \times 495)^2 = (237\,{\rm MeV})^4& \;{\rm cf.} \;& 
(206)^4 = (5.5 + 130)\times 241^3\,.
\end{eqnarray}
Hence, while it is good for the $\pi$, using the vacuum quark condensate for
the kaon leads one to overestimate $m_s^{1 {\rm GeV}}$ by 70\%!
\subsection{Bethe-Salpeter Equation}
The explicit form of the Bethe-Salpeter equation consistent with
Eq.~(\ref{ouransatz}) is 
\begin{eqnarray}
\label{bsemod}
\lefteqn{0 = \Gamma_H(k;P) \,+ }\\
& & \nonumber
\int^\Lambda\,\frac{d^4 q}{(2\pi)^4}\,
{\cal G}((k-q)^2)\, D_{\mu\nu}^{\rm free}(k-q)
 \frac{\lambda^a}{2}\gamma_\mu {\cal S}(q_+)\Gamma_H(q;P){\cal S}(q_-)
\frac{\lambda^a}{2}\gamma_\nu = 0\,.
\end{eqnarray}
With the {\it Ansatz} for ${\cal G}(k^2)$ and the solution of the quark DSE
for $S(p)$, the kernel of the BSE is specified completely.  It is a linear
integral equation and solving to obtain $\Gamma_H(k;P)$ and the bound state
mass is a straightforward numerical problem.  Having $D_{\mu\nu}(k)$, $S(p)$
and $\Gamma_H(k;P)$, the calculation of observables such as: the leptonic
decay constant; meson charge radius, $\langle \bar q q\rangle^H$; and
electromagnetic form factor, $F_H(Q^2)$; etc., is possible.

Using this exemplar one obtains, in MeV,
\begin{equation}
\label{results}
\begin{array}{lllllllllllllllll}
m_\pi &&
f_\pi &&
f^0   &&
m_{s \bar s}    &&
f_{s\bar s}  &&
m_K^{\eta_P=1/2}   &&
f_K^{\eta_P=1/2}    &&
m_K^{\eta_P=0}   &&
f_K^{\eta_P=0}    \\
138.5   &&
92.4    &&
89.8    &&
685     &&
129     &&
497     &&
109     &&
497     &&
109
\end{array}\,.
\end{equation}
$f^0$ is the leptonic decay constant of a pseudoscalar meson in the chiral
limit and ``$s \bar s\,$'' denotes a fictitious pseudoscalar bound state.  As
emphasised by Eq.~(\ref{results}), observables are independent of the
momentum partitioning parameter, $\eta_P$, when all the amplitudes in
Eq.~(\ref{genpibsa}) are retained in solving the BSE.

One aspect of the Bethe-Salpeter amplitudes, which is particularly important
in the calculation of electromagnetic form factors, is their behaviour at
large-$k^2$.  It is depicted in Fig.~\ref{figUV}, 
\begin{figure}[t]
\vspace*{-1.0em}

\centering{\
\epsfig{figure=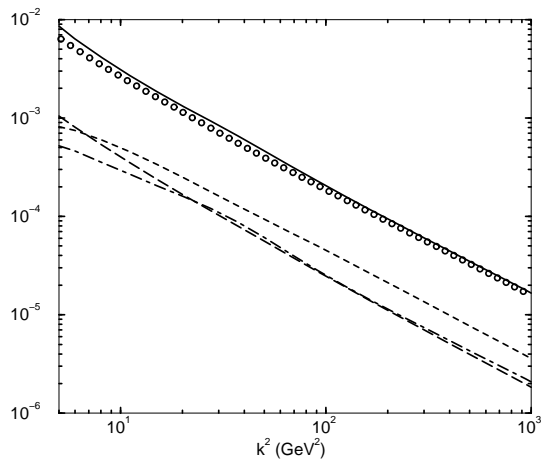,height=7.0cm}
}\vspace*{-1.5\baselineskip}
\caption{Asymptotic behaviour of the 0th Chebyshev moments of the functions
in the $\pi$-meson Bethe-Salpeter amplitude: $f_\pi\,^0\!E_\pi(k^2)$ (GeV,
solid line); $f_\pi\,^0\!F_\pi(k^2)$ (dimensionless, long-dashed line);
$k^2\,f_\pi\,^0\!G_\pi(k^2)$ (dimensionless, dashed line); and
$k^2\,f_\pi\,^0\!H_\pi(k^2)$ (GeV, dot-dashed line).  The $k^2$-dependence is
identical to that of the chiral-limit quark mass function, $M(p^2)$,
Eq.~(\protect\ref{Mchiral}).  For other pseudoscalar mesons the momentum
dependence of these functions is qualitatively the same, although the
normalising magnitude differs.
\label{figUV}}
\end{figure}
where the zeroth Chebyshev moments are
\begin{equation}
^0\!E_H(k^2) := 
        \frac{2}{\pi}\int_0^\pi\,d\beta\,\sin^2\beta\,U_0(\cos\beta)\,
                        E_H(k^2,k\cdot P;P^2)\,,
\end{equation}
$k\cdot P := \cos\beta\sqrt{k^2 P^2}$, and similarly for the other functions.
The momentum dependence of $^0\!E_\pi(k^2)$ at large-$k^2$ is identical to
that of the chiral-limit quark mass function, $M(p^2)$ in Eq.~(\ref{Mchiral})
and characterises the form of the quark-quark interaction in the ultraviolet.
Very importantly, the same is true of $^0\!F_\pi(k^2)$, $k^2\,^0\!G_\pi(k^2)$
and $k^2\,^0\!H_\pi(k^2)$, with analogous results for other mesons.  This is
critical, e.g., because it entails$\,$\cite{mrpion} that
\begin{equation}
\label{FUV}
F_\pi(q^2) \propto \frac{\alpha_{\rm S}(q^2)}{q^2}\, \frac{(-\langle \bar q
        q\rangle^0_{q^2})^2}{f_\pi^4}\,;
\end{equation}
i.e., $q^2 F_\pi(q^2) \approx {\rm const.}$, up to calculable $\ln
q^2$-corrections.  If one erroneously neglects $F$ and $G$ in $\Gamma_\pi$,
then$\,$\cite{cdrpion} $q^4 F_\pi(q^2)\approx {\rm const.}$
\section{Heavy Quarks}\label{sec:hq}
The CKM matrix characterises the difference between the mass eigenstates
($q$) and weak eigenstates ($q^\prime$) in the Standard Model:
\begin{equation}
\left(
\begin{array}{c}
d^\prime\\ s^\prime\\ b^\prime
\end{array}
\right) = \left(\begin{array}{ccc} V_{ud} & V_{us} & V_{ub} \\ V_{cd} &
V_{cs} & V_{cb} \\ V_{td} & V_{ts} & V_{tb}
\end{array}
\right) \left(
\begin{array}{c}
d\\ s\\ b
\end{array}
\right)\,,
\end{equation}
and the Standard Model requires that the CKM matrix be unitary.  The matrix
elements $V_{qQ}$ are measurable in the semileptonic decay of a pseudoscalar
meson:
\begin{equation}
A(P_{H_1} \to P_{H_2}\ell\nu) = 
\frac{G_F}{\surd 2} \,\mbox{ $V_{qQ}$} \,
\bar\ell \gamma_\mu (1 -\gamma_5)\nu\, M_\mu^{P_{H_1} P_{H_2}}(p_1,p_2)\,,
\end{equation}
where $G_F$ is the Fermi weak-decay constant and the hadronic current is
\begin{eqnarray}
M_\mu^{P_{H_1} P_{H_2}}(p_1,p_2) & := &
\langle P_{H_2}(p_2)| \bar q \gamma_\mu Q | P_{H_1}(p_1)\rangle\\
\label{fpfm}
& = & f_+(t)\, (p_1 + p_2)_\mu + f_-(t) \,q_\mu\,,
\end{eqnarray}
with $t := - q^2$.  Hence accurate measurements and calculations of these
decays can decide whether or not the Standard Model is complete.

All the information about strong interaction effects in these processes is
contained in the form factors, $f_\pm(t)$, appearing in the hadronic current,
and their accurate estimation is essential to the extraction of $V_{qQ}$ from
a measurement of a semileptonic decay rate:
\begin{eqnarray}
\label{branching}
\lefteqn{\Gamma(P_{H_1} \to P_{H_2}\ell\nu)= }\\
&& \nonumber
\frac{G_F^2}{192 \pi^3}\,|V_{qQ}|^2\,\frac{1}{m_{H_1}^3}\,
\int_0^{t_-}\,dt\,|f_+(t)|^2\,
\left[(t_+-t) (t_- - t)\right]^{3/2}\,,
\end{eqnarray}
with $t_\pm := (m_{H_1}\pm m_{H_2})^2$ and neglecting the lepton mass.  As
with all form factors, the calculation of $f_\pm(t)$ requires a knowledge of
the propagation characteristics of the mesons' constituents and the structure
of the bound state.  The DSEs are therefore ideally suited to this analysis. 

In Sec.~\ref{subsec:B2} I introduced ${\cal L}_f$, which measures whether a
quark is light or heavy.  If a quark is heavy then many simplifications ensue
in the calculation of hadronic observables, and they can all be traced to
being able to ignore the momentum dependence of that quark's self energy.
The $b$-quark is certainly heavy but one cannot be as certain about the
$c$-quark because its mass function exhibits significantly more
momentum-dependence.  Whether the $c$-quark is heavy is a qualitatively
important quantitative question.  It can be addressed by exploring the
consequences of assuming
\begin{eqnarray}
S_{b}(p) = \frac{1}{i \gamma\cdot p + \hat M_b} , &\;&
S_{c}(p) = \frac{1}{i \gamma\cdot p + \hat M_c}\,,
\end{eqnarray}
where $\hat M_f \sim M_f^E$.  I will call this the ``heavy-quark'' limit.

In this limit, when the other constituent is light, it is natural to work
with $\eta_P=1$ in Eq.~(\ref{genbse}) since the total momentum of the bound
state is dominated by the momentum of the heavy-quark.  Introducing the
heavy-meson velocity $(v_\mu)$ and binding energy $(E)$ via
\begin{equation}
\label{vel}
P:= m_H\,v_\mu, \;v^2=-1\,;\; M_H:= \hat M_Q + E\,,
\end{equation}
with $M_H$ the heavy-meson mass, one finds$\,$\cite{misha} at leading order
in $1/M_H$
\begin{eqnarray}
\label{hQ}
S(q + P)  & = &  \case{1}{2}\,\frac{1 - i \gamma\cdot v}{q\cdot v - E}\,,\\
\label{hG}
\Gamma_H(q;P)  & = &  \sqrt{M_H}\,\hat\Gamma_H(q;P) \,,
\end{eqnarray}
where the canonical normalisation condition for $\hat\Gamma_H(q;P)$ is
independent of $M_H$.  Using Eqs.~(\ref{hQ}) and (\ref{hG}) in
Eq.~(\ref{caint}) one obtains
\begin{equation}
\label{fHhQ}
f_H \propto 1/\sqrt{ M_H}\,.
\end{equation}
This is obviously not correct for light quarks since $f_\pi < f_K $, and the
experimental situation is depicted in Fig.~\ref{hQfHpic}.  It is clear from
the figure that for the $c$-quark at least the $1/\hat M_Q$-corrections to
Eq.~(\ref{fHhQ}) are significant.
\begin{figure}[t]
\centering{\ \epsfig{figure=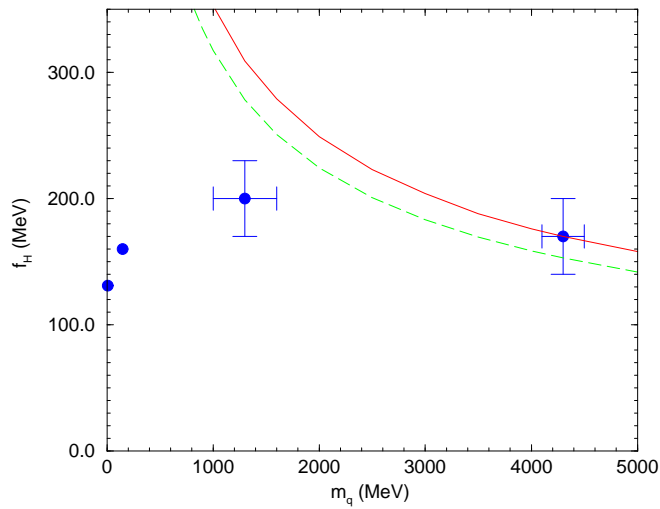,height=6.7cm} }\vspace*{-0.8em}
\caption{$f_H$ as a function of the mass of the heaviest constituent, $\hat
m_q$.\protect\cite{pdg96} (I have included an extra, multiplicative factor of
$\surd 2$, so that $f_\pi = 131\,$MeV, which is conventional in this
application.) The solid line is $f_H(\hat m_q)= {\rm const.}/\surd \hat m_q$,
fitted to $f_B$, while the dashed line assumes a 10\% shift in $f_B$ from
$1/\hat m_q$-corrections.
\label{hQfHpic}}
\end{figure}

The mass formula of Sec.~\ref{sec:MF} has another corollary in the
heavy-quark limit: using Eq.~(\ref{fHhQ}) with Eqs.~(\ref{rH}) and
(\ref{gmora}) yields
\begin{equation}
\label{mHhQ}
M_H \propto \hat m_Q\,.
\end{equation}
A model study$\,$\cite{mr98} shows this to be valid for $\hat m_Q \gsim \hat
m_s$, which is confirmed by data, Fig.~\ref{mHmq}.
\begin{figure}[t]
\centering{\ \epsfig{figure=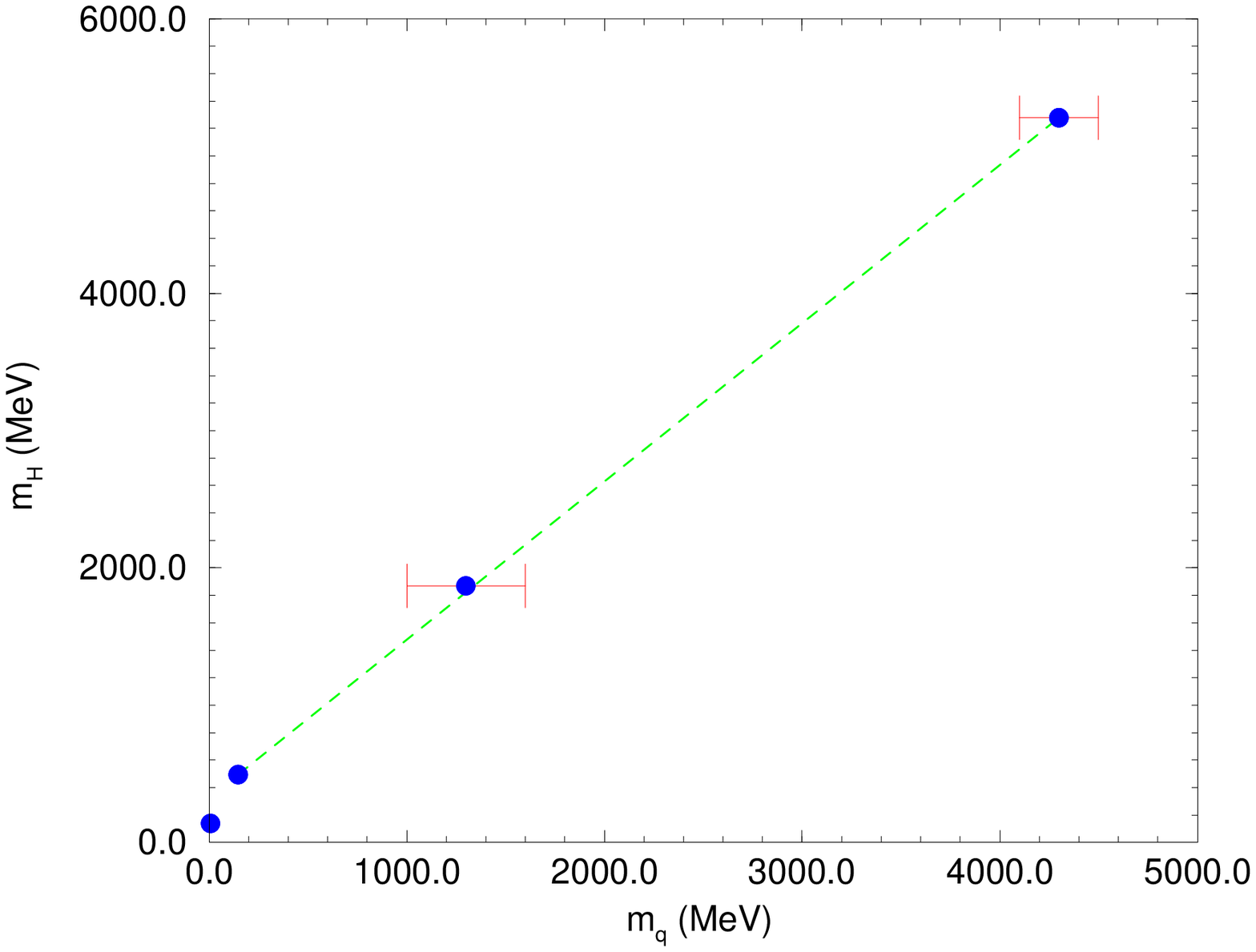,height=7.0cm} }\vspace*{-0.4\baselineskip}
\caption{Pseudoscalar meson mass as a function of the mass of the heaviest
constituent, $\hat m_q$.\protect\cite{pdg96} Only the $\pi$ does not lie on
the same straight line.  Since $\hat m_s \lsim M_\chi$, the mass-scale
associated with DCSB, it is not too surprising that Eq.~(\protect\ref{mHhQ})
is valid for $\hat m_Q \gsim \hat m_s$: that is the domain on which explicit
chiral symmetry breaking overwhelms DCSB.
\label{mHmq}}
\end{figure}
Although unsurprising when one considers that ${\rm O}\left(\hat
m_q^2\right)$-corrections to Eq.~(\ref{gmoreKp}) are large, as noted after
Eq.~(\ref{Kcf}), I observe that Eq.~(\ref{fHhQ}) is not valid until at least
$\hat m_Q \gsim \hat m_c$.  Equation~(\ref{mHhQ}) is then only valid because
of cancellations between the $1/\hat M_Q$-corrections to $f_H$ and $\langle
\bar q q \rangle^H_\zeta$.
\subsection{Semileptonic Heavy $\to$ Heavy Decays}
In impulse approximation the hadronic current describing $B_f\to D_f$ decays
is
\begin{eqnarray}
\label{ia}
\lefteqn{ M_\mu^{P_{H_1} P_{H_2}}(p_1,p_2) = 
\frac{N_c}{16\pi^4}\,
\int d^4k\,
}\\
&& \nonumber
{\rm tr}\left[
\bar\Gamma_{H_2}(k;-p_2) 
S_q(k+p_2) 
i {\cal V}_\mu^{qQ}(k+p_2,k+p_1)
S_Q(k+p_1)
\Gamma_{H_1}(k;p_1) 
S_{q^\prime}(k)\right]\,,
\end{eqnarray}
where $\bar\Gamma_{H_2}(k;-p_2)^t:= C^\dagger \Gamma_{H_2}(-k;-p_2)C,\;
C=\gamma_2\gamma_4$, $M^t$ is the matrix transpose of $M$; and ${\cal
V}_\mu^{qQ}(k_1,k_2)$ is the vector part of the dressed-quark-W-boson vertex.
Using Eqs.~(\ref{hQ}) and (\ref{hG}), and the approximation ${\cal
V}_\mu^{qQ}(k_1,k_2)= \gamma_\mu$, which is valid in the heavy-quark limit,
\begin{equation}
\label{onefunction}
f_\pm(t)  =  \case{1}{2}\, 
\frac{m_{D_f} \pm m_{B_f}}{ \sqrt{m_{D_f} m_{B_f}} }\,{ \xi_f(w)} \,;
\end{equation}
i.e., the form factors are determined by a {\it single}, {\it universal}
function: $\xi_f(w)$.\cite{iw90}

Employing an {\it Ansatz}$\;$\footnote{
Ladder-like truncations of the BSE are inadequate when the mass of one or
both constituents becomes large; e.g., this truncation does not yield the
Dirac equation when one constituent becomes infinitely massive.  Pending
improved BSE studies, an {\it Ansatz} is used.}
for the heavy-meson Bethe-Salpeter amplitude
\begin{equation}
\label{hmbsa}
\hat\Gamma_{H_{1f}}(k;p_1) = 
\gamma_5 \left(1 + \case{1}{2} i \gamma\cdot v\right)
\case{1}{\kappa_f}\,\varphi(k^2)\,,
\end{equation}
where $\kappa_f$ is the fixed, canonical Bethe-Salpeter normalisation
constant for $\hat \Gamma_{H_{1f}}$ and $\phi(k^2)$ is any simple function
that represents the heavy-meson as a finite-size, composite object; e.g.,
\begin{equation}
\label{phia}
\phi(k^2) = \exp(-k^2/\Lambda^2)\,,
\end{equation}
with $\Lambda$ a free fitting parameter, one obtains
\begin{equation}
\xi_f(w)  =  \kappa_f^2\,\frac{N_c}{32\pi^2}\,
\int_0^1 d\tau\,\frac{1}{W}\,
\int_0^\infty du \, \varphi(z_W)^2\,
        \left[\sigma_S^f(z_W) + \sqrt{\frac{u}{W}} \sigma_V^f(z_W)\right]\,,
\end{equation}
with $\sigma_{S/V}^f$ the functions describing the propagation of the
light-quark constituent, $W= 1 + 2 \tau (1-\tau) (w-1)$, $z_W= u - 2 E
\sqrt{u/W}$ and$\,$\footnote{The minimum physical value of $w$ is $w_{\rm
min}=1$, which corresponds to maximum momentum transfer with the final state
meson at rest; the maximum value is $w_{\rm max} \simeq (m_{B_f}^2 +
m_{D_f}^2)/(2 m_{B_f} m_{D_f}) = 1.6$, which corresponds to maximum recoil of
the final state meson with the charged lepton at rest.}
\begin{equation}
w = \frac{m_{B_f}^2 + m_{D_f}^2 - t}{2 m_{B_f} m_{D_f}} = \,-v_{B_f} \cdot
v_{D_f}\,. 
\end{equation}
The normalisation of the Bethe-Salpeter amplitude ensures automatically that
\begin{equation}
\label{normxi}
\xi_f(w=1) = 1\,.
\end{equation}
Experimentally measured violations of Eqs.~(\ref{onefunction}) and
(\ref{normxi}) gauge the fidelity of the heavy-quark limit for physical
processes.
\subsection{Semileptonic Heavy $\to$ Light Decays}
The heavy-quark limit is not really helpful in studying the decays $B \to
\pi$, $D\to K$ and $D\to \pi$ because there are only light-quarks in the
final state and hence no useful expansion parameter.  Therefore a theoretical
description of these decays relies heavily on a good understanding of light
quark propagation characteristics and the internal structure of light mesons.
In this case, using Eqs.~(\ref{hQ}) and (\ref{hG}), the impulse approximation
to the hadronic current yields$\,$\cite{misha}
\begin{equation}
\label{fphl}
f_+^{H_1 H_2}(t) = \kappa_{q^\prime}
         \frac{\surd 2}{f_{H_2}}\,\frac{N_c}{32\pi^2}\,
        F_{q^\prime}(t;E,m_{H_1},m_{H_2}) \,,
\end{equation}
where
\begin{eqnarray}
\lefteqn{F_{q^\prime}(t;E,m_{H_1},m_{H_2}) =}
\\ \nonumber
&&  \frac{4}{\pi}\,\int_{-1}^{1}\,\frac{d\gamma}{\sqrt{1-\gamma^2}}\,
        \int_0^1\,d\nu\,
        \int_0^\infty u^2 du\,\varphi(z_1)\,
        {\cal E}(z_1)\,W_{q^\prime}(\gamma,\nu,u) \,,
\end{eqnarray}
with ${\cal E}$ the pseudoscalar part of the light-meson Bethe-Salpeter
amplitude and
\begin{eqnarray}
W_{q^\prime}(\gamma,\nu,u) & = & 
2 \tau^2 \left[ \sigma_S^u(z_1) \frac{d}{dz_2}\sigma_V^{q^\prime}(z_2) 
       - \sigma_V^u(z_1)\frac{d}{dz_2}\sigma_S^{q^\prime}(z_2) \right]+\\
\nonumber
&  & \left(1 - \frac{u \,\nu}{m_{H_1}}\right)\,
        \sigma_S^u(z_1)\,\sigma_V^{q^\prime}(z_2) + \\
\nonumber
&  & \frac{1}{m_{H_1}}
\left[ \rule{0mm}{1.3\baselineskip}
        \sigma_S^u(z_1)\,\sigma_S^{q^\prime}(z_2)
        + u\,\nu\,\sigma_V^u(z_1)\,\sigma_S^{q^\prime}(z_2) \right.+ \\
\nonumber
&& \left.  \left(z_1 + u\,\nu\,M_{H_1}\right)
                \sigma_V^u(z_1)\,\sigma_V^{q^\prime}(z_2)
        - 2 m_{H_2}^2\,\tau^2\,\sigma_V^u(z_1)\,
                \frac{d}{dz_2}\sigma_V^{q^\prime}(z_2)\right]\,,
\end{eqnarray}
where $\sigma^\prime(x):= \frac{d}{dx}\sigma(x)$ and
\begin{eqnarray}
z_1 & = & u^2-2 u \,\nu \,E\,, \\
z_2 & = & u^2 - 2 u \,\nu\, (E- X) - m_{H_2}^2 + 
2 i \,m_{H_2} \,\gamma \,u \,\sqrt{1-\nu^2} \,,\\
\label{defX}
X & = & (m_{H_1}/2)\,[1 + (m_{H_2}^2-t)/m_{H_1}^2]\,,\\
\tau & = & u\,\sqrt{1-\nu^2}\,\sqrt{1-\gamma^2}\,.
\end{eqnarray}
Under the assumption of isospin symmetry $\sigma^u$ also represents a
$d$-quark and, to illustrate Eq.~(\ref{fphl}), the $B^0\to \pi^- \ell^+
\nu_\ell$ decay is described by
\begin{equation}
f_+^{B\pi}(t)= \kappa_d\frac{\surd 2}{f_\pi}
                \frac{N_c}{32\pi^2}\,
                F_{d}(t;E,m_B,m_\pi) \,.
\end{equation}
\subsection{Calculated Transition Form Factors}
With the dressed-light-quark propagators and light-meson Bethe-Salpeter
amplitudes fixed completely in studies$\,$\cite{brt96} of $\pi$- and
$K$-meson properties the calculation of the heavy-meson transition form
factors is straightforward.  There are two free parameters: the binding
energy, $E$, introduced in Eq.~(\ref{vel}) and the width, $\Lambda$, of the
heavy meson Bethe-Salpeter amplitude, introduced in Eq.~(\ref{phia}).  They
were fixed$\,$\cite{misha} at
\begin{equation}
E = 0.44\,{\rm GeV}\,,\;\;
\Lambda = 1.4\,{\rm GeV} = 1/[0.14\,{\rm fm}]
\end{equation}
by requiring a best, weighted least-squares fit to the three available
lattice data points$\,$\cite{latt} for $f_+^{B\pi}$ and the experimental
value$\,$\cite{cleo96} for the ${ B}^0\to\pi^- \ell^+\nu$ branching ratio.
In doing that the study used $m_B=5.27\,$GeV and was constrained to yield
$f_B = 0.17\,$GeV, which is the central value favoured in a recent analysis
of lattice simulations.\cite{hqlat} This procedure assumes only that the
$b$-quark is in the heavy-quark domain and yields $f_{B_s}=0.18\,$GeV.

The calculated form of $f_+^{B\pi}(t)$ is presented in
Fig.~\ref{figbpi}.  
\begin{figure}[t]
\vspace*{-1.0em}

\centering{\ \epsfig{figure=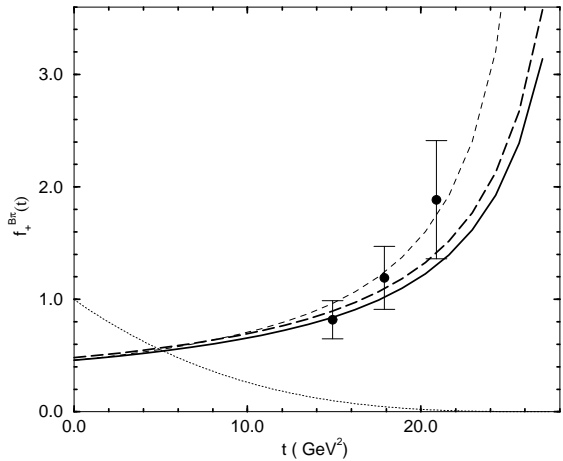,height=7.2cm}}\vspace*{-2.0em}
\caption{Calculated form of $f_+^{B\pi}(t)$.  The solid line was
obtained$\,$\protect\cite{misha} assuming only that the $b$-quark is heavy,
the dashed line assumed the same of the $c$-quark.  The data are the results
of a lattice simulation$\,$\protect\cite{latt} and the light, short-dashed
line is a vector dominance, monopole model: $f_+(t)=
0.46/(1-t/m_{B^\ast}^2)$, $m_{B^\ast} = 5.325\,$GeV.  The light, dotted line
is the phase space factor $|f_+^{B\pi}(0)|^2
\left[(t_+-t)(t_--t)\right]^{3/2}/(\pi m_B)^3$ in
Eq.~(\protect\ref{branching}), which illustrates that the $B\to \pi e \nu$
branching ratio is determined primarily by the small-$t$ behaviour
$f_+^{B\pi}(t)$.  The calculated branching ratio is $2.0\times 10^{-4}$, cf.
the experimental value$\,$\protect\cite{cleo96} $[1.8 \pm 0.4]\times
10^{-4}$.
\label{figbpi}}
\end{figure}
A good {\it interpolation} of the result is provided by
\begin{equation}
f_+^{B\pi}(t)= \frac{0.458}{1 - t/m_{\rm mon}^2}\,,
\; m_{\rm mon} = 5.67\,{\rm GeV} \,.
\end{equation}
This value of $m_{\rm mon}$ can be compared with that obtained in a fit to
lattice data:$\,$\cite{latt} $m_{\rm mon}= 5.6 \pm 0.3$.  The calculated
value of $f_+^{B\pi}(0)= 0.46$ is compared with its value obtained using a
range of other theoretical tools in Table~\ref{comp}.
\begin{table}[h,t]
\begin{center}
\begin{tabular}{l|l}
 Reference              & $f_+^{B\pi}(0) $    \\\hline
Ref.~[\protect\ref{mishaR}]  & 0.46  \\\hline
Dispersion relations~\cite{lellouch96} & 0.18 $\to$ 0.49 \\
Quark Model~\cite{qma}  & $ 0.33 \pm 0.06 $ \\
Quark Model~\cite{qmb}  & $0.21 \pm 0.02 $ \\
Quark Model~\cite{qmc}  & 0.29 \\
Light-Cone Sum Rules~\cite{lcsr} & $\!\!\left\{\begin{array}{l}
                                   0.29~{\rm direct}\\  
                                   0.44~{\rm pole~dominance}
                                        \end{array} \right.$\\
Quark Confinement Model~\cite{mishaB}
                        & 0.6 \\
Quark Confinement Model~\cite{mishaC,mishaD} & 0.53 
\end{tabular}
\end{center}
\caption{Comparison of the calculated result for $f_+^{B\pi}(0)$ with a
values obtained using other theoretical tools.  More extensive and
complementary lists are presented
elsewhere.\protect\cite{lellouch96,qmc,mishaD}
\label{comp}}
\end{table}
The uncertainty in $f_+^{B\pi}(0)$ is the major source of the large
uncertainty in the CKM matrix element: $V_{bu}= 0.0032 \pm 0.0012$.

The calculated form of $\xi(w)$, which characterises the semileptonic heavy-
$\to$ heavy-meson decays, is depicted in Fig.~\ref{figiwfn}.
\begin{figure}[t]
\vspace*{-1.0em}

\centering{\
\epsfig{figure=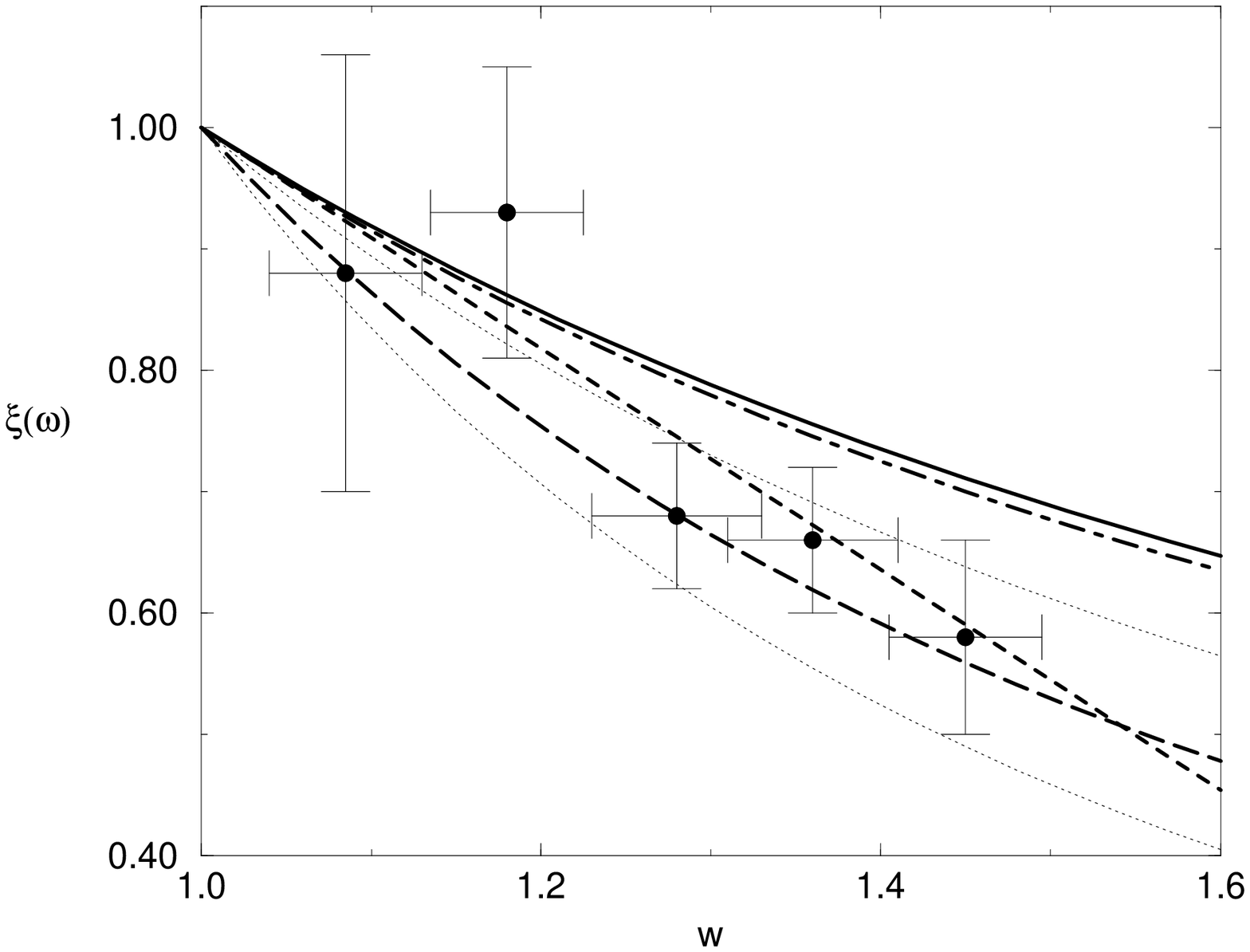,height=7.0cm}}\vspace*{-2.0em}
\caption{Calculated form of $\xi(w)$ cf. recent experimental analyses.  The
solid line was obtained$\,$\protect\cite{misha} assuming only that the
$b$-quark is heavy, the dash-dot line assumed the same of the $c$-quark.
Experiment: data points - Ref.~[\protect\ref{argus93R}]; short-dashed line -
linear fit from Ref.~[\protect\ref{cesr96R}],
$\xi(w)  =  1 - \rho^2\,( w - 1), \; \rho^2 = 0.91\pm 0.15 \pm 0.16\,$;
long-dashed line - nonlinear fit from Ref.~[\protect\ref{cesr96R}],
$\xi(w)  =  [2/(w+1)]\,\exp\left[(1-2\rho^2) \,(w-1)/(w+1)\right],
        \;\rho^2 = 1.53 \pm 0.36 \pm 0.14\,$.
The two light, dotted lines are this nonlinear fit evaluated with the extreme
values of $\rho^2$: upper line, $\rho^2= 1.17$ and lower line, $\rho^2=1.89$.
\label{figiwfn}}
\end{figure}
It yields a value for $\rho^2:= - \xi^\prime(w=1)= 0.87-0.92$, close to that
obtained with a linear fitting form,\cite{cesr96} however, $\xi(w)$ has
significant curvature and deviates quickly from that fit.  The curvature is,
in fact, very well matched to that of the nonlinear fit,\cite{cesr96}
however, the value of $\rho^2$ reported in that case is very different from
the calculated value.  The derivation of the formula for $\xi(w)$ assumes
that the heavy-quark limit is valid not only for the $b$-quark but also for
the $c$-quark.  These results therefore suggest that the latter assumption is
only accurate to approximately 20\%.

The calculated form of $f_+^{DK}(t)$ is depicted in Fig.~\ref{figdk}.
\begin{figure}[t]
\centering{\
\epsfig{figure=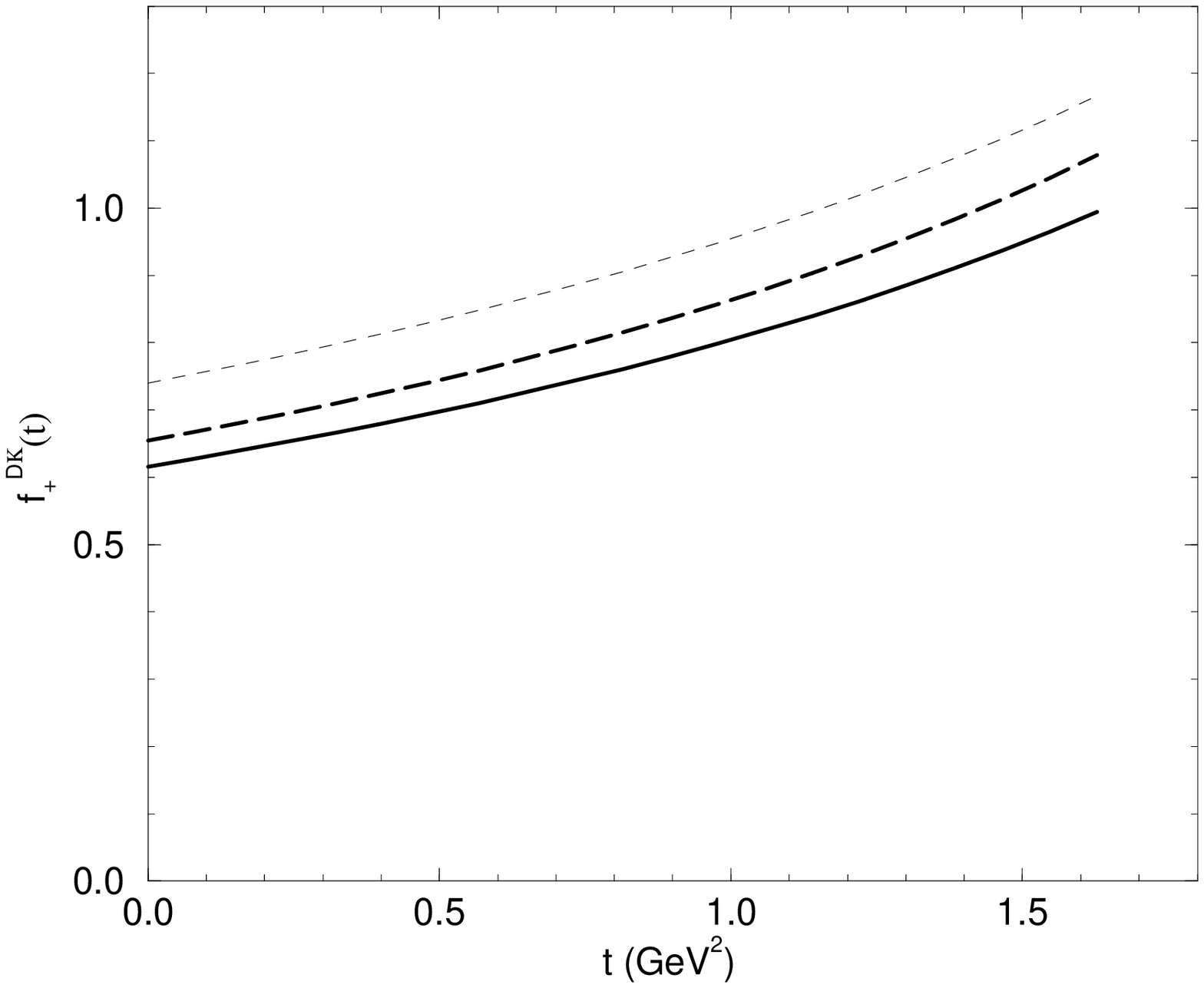,height=7.2cm}}\vspace*{-1.0em}
\caption{Calculated form of $f_+^{DK}(q^2)$: the solid line was
obtained$\,$\protect\cite{misha} assuming only that the $b$-quark is heavy,
the dashed line assumed the same of the $c$-quark.  The light, short-dashed
line is a vector dominance, monopole model: $f_+(q^2)=
0.74/(1-q^2/m_{D_s^\ast}^2)$, $m_{D_s^\ast} = 2.11\,$GeV.
\label{figdk}}
\end{figure}
The $t$-dependence is well-approximated by a monopole fit.  The calculated
value of $f_+^{DK}(0) = 0.62$ is approximately 15\% less than the
experimental value.\cite{pdg96} That is also a gauge of the size of $1/\hat
M_c$-corrections, which are expected to reduce the value of the $D$- and
$D_s$-meson leptonic decay constants calculated in the heavy-quark limit:
$f_D= 285\,$MeV, $f_{D_s}= 298\,$MeV.  A 15\% reduction yields $f_D =
0.24\,$GeV and $f_{D_s}=0.26\,$GeV, values which are consistent with lattice
estimates,\cite{hqlat} the latter with experiment,\cite{expfds} and both with
the scaling law illustrated in Fig.~\ref{hQfHpic}.

On the kinematically accessible domain, $0 < t < (m_D - m_\pi)^2$, the
calculated form of $f_+^{D\pi}(t)$ is well described by the monopole fit 
\begin{equation}
f_+^{D\pi}(t) = \frac{0.716}{1 - t/m_{\rm mon}^2}\,,
\; m_{\rm mon} = 2.15\,{\rm GeV} \,.
\end{equation}
A naive vector meson dominance assumption would employ $m_{\rm mon} =
m_{D^*}= 2.0\,{\rm GeV}\,$.  Using$\,$\cite{misha} $|V_{cd}/V_{cs}|^2 = 0.051
\pm 0.002$,\cite{pdg96} 
\begin{equation}
R_\pi := \frac{Br(D \to \pi \ell \nu)}{Br(D\to K \ell \nu)}
= 2.47 \left|\frac{V_{cd}}{V_{cs}}\right|^2 = 0.13\,.
\end{equation}
The bulk of the $1/\hat M_c$-corrections cancel in this ratio.
Experi\-ment\-ally$\,$\cite{pdg96,mark3,cleoii}
\begin{eqnarray}
\label{rpia}
R_\pi = \frac{Br(D^0\to \pi^- e^+ \nu_e)}{Br(D^0\to K^- e^+ \nu_e)} & = &
        0.11 ^{+0.06}_{-0.03} \pm 0.1\,,\\
\label{rpib}
R_\pi = 2 \frac{Br(D^+\to \pi^0 e^+ \nu_e)}{Br(D^+\to \bar K^0 e^+ \nu_e)} &
        = & 0.17 \pm 0.05 \pm 0.03~\,.
\end{eqnarray}
If one assumes single-pole, $D^*$ and $D_s^*$ vector meson dominance for the
$t$-dependence of the form factors $f_+^{D\pi}$ and $f_+^{DK}$, one obtains
the simple formula
\begin{equation}
R_\pi= 1.97 \left|\frac{f_+^{D\pi}(0)}{f_+^{DK}(0)}\right|^2
                \left|\frac{V_{cd}}{V_{cs}}\right|^2\,.
\end{equation}
That approach was used$\,$\cite{pdg96} to estimate $f_+^{D\pi}(0)/f_+^{DK}(0)
= 1.0^{+0.3}_{-0.2}\pm 0.04$ or $1.3\pm0.2\pm 0.1$ from Eqs.~(\ref{rpia}) and
(\ref{rpib}).  This calculation$\,$\protect\cite{misha} yields
\begin{equation}
\frac{f_+^{D\pi}(0)}{f_+^{DK}(0)} = 1.16\,.
\end{equation}

It must be noted that Ref.~[\ref{mishaR}] explicitly {\it did not}\ assume
vector meson dominance.  The calculated results reflect only the importance
and influence of the dressed-quark and -gluon substructure of the heavy
mesons.  That substructure is manifest in the dressed propagators and bound
state amplitudes, which fully determine the value of every calculated
quantity.  Explicit vector meson contributions would appear as pole terms in
${\cal V}_\mu^{f_1f_2}(k_1,k_2)$, which are excluded in the {\it Ansatz}:
${\cal V}_\mu^{qQ}(k_1,k_2)= \gamma_\mu$.  That simple-pole {\it Ans\"atze}
provide efficacious interpolations of the calculated results on the
accessible kinematic domain is not surprising, given that the form factor
must rise slowly away from its value at $t=0$ and the heavy meson mass
provides a dominant intrinsic scale, which is only modified slightly by the
scale in the light-quark propagators and meson bound state amplitudes.
Similar observations are true in the calculation of the pion form 
factor.\cite{cdrpion,pctrev}

In this section I illustrated a heavy-quark limit of the DSEs, based on the
observation that the mass function of heavy-quarks evolves slowly with
momentum, and the manner in which it can be used to describe the properties
of heavy-mesons.  They are much like light-mesons: bound states of finite
extent, with dressed-quark constituents.  The analysis of $B\to D$, $D\to K$
and $D\to \pi$ transitions indicates that the heavy-quark limit is accurate
to within 15-20\% for the $c$-quark.  A significant feature is the
correlation of heavy $\to$ heavy and heavy $\to$ light transitions {\it and}
their correlation with light meson observables, which are dominated by
effects such as confinement and DCSB.
\section{Finite $T$ and $\mu$}
Nonperturbative methods are necessary to study the transition to a QGP, which
is characterised by qualitative changes in order parameters such as the quark
condensate.  One widely used approach is the numerical simulation of
finite-$T$ lattice-QCD, with the first simulations in the early eighties and
extensive efforts since then.\cite{karsch95} They have contributed
considerably to the current understanding of the nature of the QGP.  The
commonly used quenched approximation is inadequate for studying the phase
diagram of finite-$T$ QCD because the details of the transition depend
sensitively on the number of active (light) flavours.  It is therefore
necessary to include the fermion determinant.

That is even more important in the presence of $\mu$, which modifies the
fermion piece of the Euclidean action: $\gamma\cdot \partial + m \to
\gamma\cdot \partial - \gamma_4 \mu + m$, introducing an explicit imaginary
part to the fermion determinant.  With the $\mu\neq 0$ lattice-QCD action
being complex, the study of finite density is significantly more difficult
than that of finite temperature.  Simulations that ignore the fermion
determinant encounter a forbidden region, which begins at $\mu =
m_\pi/2$,\cite{dks96} and since $m_\pi\to 0$ in the chiral limit this is a
serious limitation, preventing a reliable study of chiral symmetry
restoration.  The phase of the fermion determinant is essential in
eliminating this artefact.\cite{adam}

The contemporary application of DSEs at finite temperature and chemical
potential is a straightforward extension of the $T=0=\mu$
studies.\cite{iitap} The direct approach is to develop a finite-$T$ extension
of {\it Ans\"atze} for the dressed-gluon propagator.  The quark DSE can then
be solved and, having the dressed-quark and -gluon propagators, the response
of bound states to increases in $T$ and $\mu$ can be calculated.  As a
nonperturbative approach that allows the simultaneous study of confinement
and DCSB, the DSEs have a significant overlap with lattice simulations: each
quantity that can be estimated using lattice simulations can also be
calculated using DSEs.  That means they can be used to check the lattice
simulations, and importantly, that lattice simulations can be used to
constrain their model-dependent aspects.  Once agreement is obtained on the
common domain, the DSEs can be used to explore phenomena presently
inaccessible to lattice simulations.
\subsection{Finite-$(T,\mu)$ Quark DSE}
The renormalised dressed-quark propagator at finite-$(T,\mu)$ has the form
\begin{eqnarray}
\label{genformS}
S(\vec{p},\tilde\omega_k)  & = & \frac{1}
{i\vec{\gamma}\cdot \vec{p}\,A(\vec{p},\tilde\omega_k)
+i\gamma_4\,\tilde\omega_k C(\vec{p},\tilde\omega_k)+
B(\vec{p},\tilde\omega_k)}\\
& \equiv&  -i\vec{\gamma}\cdot \vec{p}\,\sigma_A(\vec{p},\tilde\omega_k)
-i\gamma_4\,\tilde\omega_k \sigma_C(\vec{p},\tilde\omega_k)+
\sigma_B(\vec{p},\tilde\omega_k)\,,
\end{eqnarray}
where $\tilde\omega_k := \omega_k + i \mu $ with $\omega_k= (2 k + 1)\,\pi
\,T$, $k\in {\rm Z}\!\!\!{\rm Z}$, the fermion Matsubara frequencies.  The
complex scalar functions: $A(\vec{p},\tilde\omega_k)$,
$B(\vec{p},\tilde\omega_k)$ and $C(\vec{p},\tilde\omega_k)$ satisfy:
$ {\cal F}(\vec{p},\tilde\omega_k)^\ast = {\cal
F}(\vec{p},\tilde\omega_{-k-1})\,, $
${\cal F}=A,B,C$, and although not explicitly indicated they are functions
only of $|\vec{p}|^2$ and $\tilde\omega_k^2$.  Using the
Matsubara$\,$\cite{rivers} (or imaginary-time) formalism $O(4)$ covariance,
the Euclidean realisation of Poincar\'e covariance, is broken to $O(3)$ and
only systems in equilibrium can be studied.

$S(\vec{p},\tilde\omega_k)$ satisfies the DSE
\begin{equation}
\label{qDSE}
S^{-1}(\vec{p},\tilde\omega_k) = Z_2^A \,i\vec{\gamma}\cdot \vec{p} + Z_2 \,
(i\gamma_4\,\tilde\omega_k + m_{\rm bm})\, 
        + \Sigma^\prime(\vec{p},\tilde\omega_k)\,,
\end{equation}
where the regularised self energy is
\begin{eqnarray}
\Sigma^\prime(\vec{p},\tilde\omega_k) & = & i\vec{\gamma}\cdot \vec{p}
\,\Sigma_A^\prime(\vec{p},\tilde\omega_k) + i\gamma_4\,\tilde\omega_k
\,\Sigma_C^\prime(\vec{p},\tilde\omega_k) +
\Sigma_B^\prime(\vec{p},\tilde\omega_k)\; ,
\end{eqnarray}
\begin{equation}
\Sigma_{\cal F}^\prime(\vec{p},\tilde\omega_k)  = 
\int_{l,q}^{\bar\Lambda}\, \case{4}{3}\,g^2\,
D_{\mu\nu}(\vec{p}-\vec{q},\tilde\omega_k-\tilde\omega_l)\case{1}{4} {\rm
tr}\left[{\cal P}_{\cal F} \gamma_\mu
S(\vec{q},\tilde\omega_l)
\Gamma_\nu(\vec{q},\tilde\omega_l;\vec{p},\tilde\omega_k)\right]\,,
\label{regself}
\end{equation}
$\int_{l,q}^{\bar\Lambda}:=\, T
\,\sum_{l=-\infty}^\infty\,\int^{\bar\Lambda}\frac{d^3q}{(2\pi)^3}$ and
${\cal P}_A:= -(Z_1^A/p^2)i\gamma\cdot p$, ${\cal P}_B:= Z_1 $, ${\cal P}_C:=
-(Z_1/\tilde\omega_k)i\gamma_4$.  The finite-$(T,\mu)$, Landau-gauge
dressed-gluon propagator in the kernel has the form
\begin{eqnarray}
g^2 D_{\mu\nu}(\vec{p},\Omega) & = &
P_{\mu\nu}^L(\vec{p},\Omega) \,\Delta_F(\vec{p},\Omega) + 
P_{\mu\nu}^T(\vec{p})\, \Delta_G(p,\Omega) \,,\\
P_{\mu\nu}^T(\vec{p}) & := &\left\{
\begin{array}{ll}
0; \; & \mu\;{\rm and/or} \;\nu = 4,\\
\displaystyle
\delta_{ij} - \frac{p_i p_j}{|\vec{p}|^2}; \; & \mu,\nu=i,j=1,2,3
\end{array}\right.\,,
\end{eqnarray}
with $P_{\mu\nu}^T(p) + P_{\mu\nu}^L(p,p_4) = \delta_{\mu\nu}- p_\mu
p_\nu/({\sum_{\alpha=1}^4 \,p_\alpha p_\alpha})$; $\mu,\nu= 1,\ldots, 4$.  

Whether or not $(T,\mu)$ are nonzero, in studying confinement one cannot
assume that the analytic structure of a dressed propagator is the same as
that of the free particle propagator: it must be determined dynamically.  The
$\tilde p_k:=(\vec{p},\tilde\omega_k)$-dependence of $A$ and $C$ is
qualitatively important since it can conspire with that of $B$ to eliminate
free-particle poles in the dressed-quark propagator.\cite{burden} In that
case the propagator does not have a Lehmann representation so that, in
general, the Matsubara sum cannot be evaluated analytically.  More
importantly, it further complicates a real-time formulation of the finite
temperature theory making the study of nonequilibrium thermodynamics a very
challenging problem.  In addition, the $\tilde p_k$-dependence of $A$ and $C$
can be a crucial factor in determining the behaviour of bulk thermodynamic
quantities such as the pressure and entropy, being responsible for these
quantities reaching their respective ultrarelativistic limits only for very
large values of $T$ and $\mu$.  It is therefore important in any DSE study to
retain $A(\tilde p_k)$ and $C(\tilde p_k)$, and their dependence on $\tilde
p_k$.
\subsection{Phase Transitions and Order Parameters}
One order parameter for the chiral symmetry restoration transition is the
quark condensate, defined via the renormalised dressed-quark propagator,
Eq.~(\ref{cbqbq}).  An equivalent order parameter is
\begin{equation}
\label{chiorder}
{\cal X} := {\sf Re}\,B_0(\vec{p}=0,\tilde \omega_0)\,,
\end{equation}
which makes it clear that the zeroth Matsubara mode determines the character
of the chiral phase transition.  An order parameter for confinement, valid
for both light- and heavy-quarks, was introduced in Ref.~[\ref{prla}].  It is
a single, quantitative measure of whether or not a Schwinger function has a
Lehmann representation, and has been used$\,$\cite{m95} to striking effect in
QED$_3$.
\subsection{Study at $(T\neq 0,\mu=0)$}\label{subsec:D1}
Deconfinement and chiral symmetry restoration at $T\neq 0$ have been
studied$\,$\cite{prl} in a DSE-model of two-light-flavour QCD.  The quark DSE
was solved using a one-parameter model dressed-gluon propagator, which
provides a good description of $\pi$ and $\rho$-meson observables at
$T=0=\mu$.\cite{fr96} The transitions are coincident and second-order at a
critical temperature of $T_c\approx 150\,$MeV, with the same estimated
critical exponents: $\beta=0.33\pm 0.03$.  Both $m_\pi$ and $f_\pi$ are
insensitive to $T$ for $T\lsim 0.7\,T_c$.  The evolution with $T$ is so slow
that even at $T=0.9\,T_c$ there is only a 20\% suppression of $\Gamma_{\pi\to
\ell\nu}$.  However, for $T$ very near to $T_c$ the pion mass increases
substantially, as thermal fluctuations overwhelm quark-antiquark attraction
in the pseudoscalar channel, until, at $T=T_c$, $f_\pi\to 0$ and there is no
bound state.  These results confirm$\,$\footnote{The lattice and DSE
estimates of $\beta_{\cal X}$, however, do not survive more exhaustive
studies,\protect\cite{laermann,arne} and the most recent
analyses\protect\cite{arne,mPrivate} suggest that in DSE models whose
long-range part is described by a $\delta$-function singularity the chiral
symmetry restoration transition at finite-$T$ is described by a mean-field
value of $\beta$.}  those of contemporary numerical simulations of finite-$T$
lattice-QCD.\cite{karsch95}
\subsection{Complementary study at $(T= 0,\mu\neq 0)$}\label{subsec:D2}
The behaviour of this model at $\mu\neq 0$ has also been
explored.$\,$\cite{greg} Using the simple dressed-gluon
propagator$\,$\cite{fr96}
\begin{eqnarray}
\label{gksquare}
\frac{{\cal G}(k^2)}{k^2} & = &
\case{16}{9} \pi^2 \left[ 4 \pi^2 m_t^2 \delta^4(k)
+ \frac{1- {\rm e}^{-[k^2/(4 m_t^2)]}}{k^2}\right]\,,
\end{eqnarray}
where $m_t=0.69\;$GeV$\,= 1/[0.29\,{\rm fm}]$ is a mass-scale that marks the
boundary between the perturbative and nonperturbative domains, the quark DSE
was solved in rainbow approximation.  The solution has the form
\begin{equation}
S(p_{[\mu]}):= -i
\vec{\gamma}\cdot \vec{p}\, \sigma_A(p_{[\mu]} ) 
        - i \gamma_4\, \omega_{[\mu]} \,
\sigma_C(p_{[\mu]}) + \sigma_B(p_{[\mu]})\,, 
\end{equation}
where $p_{[\mu]}:= (\vec{p},\omega_{[\mu]})$, with $\omega_{[\mu]} := p_4 + i
\mu$.

As elucidated in Sec.~\ref{sec:sim}, there are two distinct types of
solution: a Nambu-Goldstone mode with confinement and DCSB characterised by
$\sigma_{B_0} \not\equiv 0$; and a deconfined, chirally-symmetric Wigner-Weyl
mode characterised by $\sigma_{B_0} \equiv 0$.  The possibility of a phase
transition between the two modes is explored by calculating the relative
stability of the different phases, which is measured by the difference
between their tree-level auxiliary-field effective-actions:
\begin{eqnarray}
\label{bagpres}
\lefteqn{\case{1}{2 N_f N_c}\,{\cal B}(\mu)  := 
\int^\Lambda\frac{d^4p}{(2\pi)^4}\, }\\
&& 
\nonumber 
\left\{ \ln\left[\frac{|\vec{p}|^2 A_0^2 + \omega_{[\mu]}^2 C_0^2 + B_0^2}
                {|\vec{p}|^2 \hat A_0^2 + \omega_{[\mu]}^2 \hat C_0^2}\right]
+  |\vec{p}|^2 \left(\sigma_{A_0} - \hat\sigma_{A_0}\right)
+  \omega_{[\mu]}^2 \left(\sigma_{C_0} - \hat\sigma_{C_0}\right)\right\}\,,
\end{eqnarray}
where $\hat A$ and $\hat C$ represent the solution of Eq.~(\ref{qDSE})
obtained when $B_0\equiv 0$; i.e., when DCSB is absent.  This solution exists
for all $\mu$.  ${\cal B}(\mu)$ defines a $\mu$-dependent ``bag
constant''.\cite{reg85} It is positive when the Nambu-Goldstone phase is
dynamically favoured; i.e., has the highest pressure, and becomes negative
when the Wigner pressure becomes larger.  Hence the critical chemical
potential is the zero of ${\cal B}(\mu)$, which is $\mu_c=375\,$MeV.  The
abrupt switch from the Nambu-Goldstone to the Wigner-Weyl mode signals a
first order transition.

The chiral order parameter {\it increases} with increasing chemical potential
up to $\mu_c $, with ${\cal X}(\mu_c)/{\cal X}(0)\approx 1.2$, whereas the
deconfinement order parameter, $\kappa(\mu)$, is insensitive to increasing
$\mu$.  At $\mu_c$ they both drop immediately and discontinuously to zero, as
expected of coincident, first-order phase transitions.  The increase of
${\cal X}$ with $\mu$ is a necessary consequence of the momentum dependence
of the scalar piece of the quark self energy, $B(p_{[\mu]})$.\cite{thermo}
The vacuum quark condensate behaves in qualitatively the same manner as
${\cal X}$.

Even though the chiral order parameter {\it increases} with $\mu$, $m_\pi$
{\it decreases} slowly as $\mu$ increases, with $m_\pi(\mu\approx
0.7\,\mu_c)/m_\pi(0) \approx 0.94$.  At this point $m_\pi$ begins to increase
although, for $\mu<\mu_c$, $m_\pi(\mu)$ does not exceed $m_\pi(0)$, which
precludes pion condensation.  The behaviour of $m_\pi$ results from mutually
compensating increases in $f_\pi^2$ and $\langle m_R^\zeta (\bar q
q)_\zeta\rangle_\pi$.  $f_\pi$ is insensitive to the chemical potential until
$\mu\approx 0.7\,\mu_c$, when it increases sharply so that
$f_\pi(\mu_c^-)/f_\pi(\mu=0) \approx 1.25$.  The relative insensitivity of
$m_\pi$ and $f_\pi$ to changes in $\mu$, until very near $\mu_c$, mirrors the
behaviour of these observables at finite-$T$.\cite{prl} For example, it leads
only to a $14$\% change in $\Gamma_{\pi\to \mu\nu}$ at $\mu\approx
0.9\,\mu_c$: an {\it increase} in this case.  The universal scaling
conjecture of Ref.~[\ref{brownR}] is inconsistent with the anticorrelation
observed between the $\mu$-dependence of $f_\pi$ and $m_\pi$.
\vspace*{0.5em}

\hspace*{-\parindent}{\it \arabic{section}.\arabic{subsection}.1}:~{\bf
$\mu$-dependence anticorrelated with $T$-dependence.}~Comparing the
$\mu$-dependence of $f_\pi$ and $m_\pi$ with their $T$-dependence, one
observes an anticorrelation; e.g., at $\mu=0$, $f_\pi$ falls continuously to
zero as $T$ is increased towards $T_c \approx 150\,$MeV.  This is a necessary
consequence of the momentum-dependence of the quark self energy.  In
calculating these observables one obtains expressions for $m_\pi^2$ or
$f_\pi^2$ and thus the natural dimension is mass-squared.  Therefore their
behaviour at finite $T$ and $\mu$ is determined by
\begin{equation}
{\sf Re}(\omega_{[\mu]}^2) \sim [\pi^2 T^2 - \mu^2]\,,
\end{equation}
where the $T$-dependence arises from the introduction of the fermion
Matsubara frequency: \mbox{$p_4 \to (2k +1)\pi T$}.  Hence when such a
quantity decreases with $T$ it will increase with $\mu$, and
vice-versa.\cite{schmidt98} 
\vspace*{0.5em}

\hspace*{-\parindent}{\it \arabic{section}.\arabic{subsection}.2}:~{\bf
$(-\langle \bar q q \rangle)$ increases with $\mu$.}~The confined-quark
vacuum consists of quark-antiquark pairs correlated in a scalar condensate
and increasing $\mu$ increases the scalar density: $(-\langle \bar q q
\rangle)$.  This is an expected consequence of confinement, which entails
that each additional quark must be locally paired with an antiquark thereby
increasing the density of condensate pairs as $\mu$ is increased.  For this
reason, as long as $\mu<\mu_c$, there is no excess of particles over
antiparticles {\it in the vacuum} and hence the baryon number density remains
zero;\cite{thermo} i.e., $ \rho_B^{u+d}=0\,,\; \mu < \mu_c $.  This is just
the statement that quark-antiquark pairs confined in the condensate do not
contribute to the baryon number density.
\vspace*{0.5em}

\hspace*{-\parindent}{\it \arabic{section}.\arabic{subsection}.3}:~{\bf The
core of neutron stars.}~The vacuum quark pressure, $P^{u+d}[\mu]$, can be
calculated.\cite{thermo} After deconfinement it increases rapidly, as the
condensate ``breaks-up'', and an excess of quarks over antiquarks develops.
The baryon-number density, \mbox{$\rho_B^{u+d} = (1/3)\partial
P^{u+d}/\partial \mu$}, also increases rapidly, with
\begin{equation}
\rho_B^{u+d}(\mu\approx 2 \mu_c) \simeq 3 \,\rho_0\,,
\end{equation}
where $\rho_0=0.16\,{\rm fm}^{-3}$ is the equilibrium density of nuclear
matter.  For comparison, the central core density expected in a
$1.4\,M_\odot$ neutron star is $3.6$-$4.1\,\rho_0$.\cite{wiringa}
Finally,
at $\mu\sim 5 \mu_c$, the quark pressure saturates the ultrarelativistic
limit: $P^{u+d}= \mu^4/(2\pi^2)$, and there is a simple relation between
baryon-density and chemical-potential:
\begin{equation}
\label{fqg}
\rho_B^{u_F+d_F}(\mu) = \frac{1}{3} \, \frac{2 \mu^3}{\pi^2}\,,
\; \forall \mu \gsim 5 \mu_c \,,
\end{equation}
so that $\rho_B^{u_F+d_F}(5\mu_c)\sim 350\,\rho_0$.  Thus the quark pressure
in the deconfined domain overwhelms any finite, additive contribution of
hadrons to the equation of state.  That was anticipated in Ref.~[\ref{gregR}]
where the hadron contribution was neglected.  This discussion suggests that a
QGP is likely to exist in the core of dense neutron stars.
\subsection{Simultaneous study of $(T\neq 0,\mu\neq 0)$}\label{sec:sim}
This is a difficult problem and the most complete study$\,$\cite{thermo} to
date employs a simple {\it Ansatz} for the dressed-gluon propagator that
exhibits the infrared enhancement suggested by Ref.~[\ref{bp89R}]:
\begin{equation}
\label{mnprop}
g^2 D_{\mu\nu}(\vec{p},\Omega_k) = 
\left(\delta_{\mu\nu} 
- \frac{p_\mu p_\nu}{|\vec{p}|^2+ \Omega_k^2} \right)
2 \pi^3 \,\frac{\eta^2}{T}\, \delta_{k0}\, \delta^3(\vec{p})\,,
\end{equation}
with $\Omega_k=2 k \pi T$, the boson Matsubara frequency.  As an
infrared-dominant model that does not represent well the behaviour of
$D_{\mu\nu}(\vec{p},\Omega_k)$ away from \mbox{$|\vec{p}|^2+ \Omega_k^2
\approx 0$}, some model-dependent artefacts arise.  However, there is
significant merit in its simplicity and, since the artefacts are easily
identified, the model remains useful as a means of elucidating many of the
qualitative features of more sophisticated {\it Ans\"atze}.

With this model, using the rainbow approximation, the quark DSE
is$\,$\cite{bender96}
\begin{equation}
\label{mndse}
S^{-1}(\vec{p},\omega_k) = S_0^{-1}(\vec{p},\tilde \omega_k)
        + \case{1}{4}\eta^2\gamma_\nu S(\vec{p},\tilde \omega_k) \gamma_\nu\,.
\end{equation}
A simplicity inherent in Eq.~(\ref{mnprop}) is now apparent: it allows the
reduction of an integral equation to an algebraic equation, in whose solution
many of the qualitative features of more sophisticated models are manifest.

In the chiral limit Eq.~(\ref{mndse}) reduces to a quadratic equation for
$B(\tilde p_k)$, which has two qualitatively distinct solutions.  The
Nambu-Goldstone solution, with
\begin{eqnarray}
\label{ngsoln}
B(\tilde p_k) & = &\left\{
\begin{array}{lcl}
\sqrt{\eta^2 - 4 \tilde p_k^2}\,, & &{\sf Re}(\tilde p_k^2)<\case{\eta^2}{4}\\
0\,, & & {\rm otherwise}
\end{array}\,,\right.\\
C(\tilde p_k) & = &\left\{
\begin{array}{lcl}
2\,, & & {\sf Re}(\tilde p_k^2)<\case{\eta^2}{4}\\
\case{1}{2}\left( 1 + \sqrt{1 + \case{2 \eta^2}{\tilde p_k^2}}\right)
\,,& & {\rm otherwise}
\end{array}\,,\right.
\end{eqnarray}
describes a phase of this model in which: 1) chiral symmetry is dynamically
broken, because one has a nonzero quark mass-function, $B(\tilde p_k)$, in
the absence of a current-quark mass; and 2) the dressed-quarks are confined,
because the propagator described by these functions does not have a Lehmann
representation.  The alternative Wigner solution, for which
\begin{eqnarray}
\label{wsoln}
\hat B(\tilde p_k)  \equiv  0 &,\;& 
\hat C(\tilde p_k)  = 
\case{1}{2}\left( 1 + \sqrt{1 + \case{2 \eta^2}{\tilde p_k^2}}\right)\,,
\end{eqnarray}
describes a phase of the model with neither DCSB nor confinement.  Here the
relative stability of the different phases is measured by a
$(T,\mu)$-dependent vacuum pressure difference; i.e., a $(T,\mu)$-dependent
bag constant: ${\cal B}(T,\mu)$.
 
${\cal B}(T,\mu)=0$ defines the phase boundary, and the deconfinement and
chiral symmetry restoration transitions are coincident.  For $\mu=0$ the
transition is second order and the critical temperature is $T_c^0 =
0.159\,\eta$, which using the value of $\eta=1.06\,$GeV obtained by fitting
the $\pi$ and $\rho$ masses corresponds to $T_c^0 = 0.170\,$GeV.  This is
only 12\% larger than the value reported in Sec.~\ref{subsec:D1}, and the
order of the transition is the same.  For any $\mu \neq 0$ the transition is
first-order.  For $T=0$ the critical chemical potential is
$\mu_c^0=0.3\,$GeV, which is \mbox{$\approx 30$\%} smaller than the result in
Sec.~\ref{subsec:D2}.  The discontinuity in the order parameter vanishes at
$\mu=0$, and this defines a tricritical point.

In this model the quark pressure, $P_q$, is calculated easily.  Confinement
means that $P_q \equiv 0$ in the confined domain.  In the deconfined domain
it approaches the ultrarelativistic, free particle limit, $P_{\rm UR}$, at
large values of $ T$ and $\mu$.  However, the approach to this limit is slow.
For example, at $ T \sim 0.3\,\eta \sim 2 T_c^0$, or $ \mu \sim \eta \sim 3
\mu_c^0$, $P_q$ is only $0.5\,P_{\rm UR}$.  A qualitatively similar result is
observed in numerical simulations of finite-$T$ lattice-QCD.\cite{karsch95}
This feature results from the persistence of momentum dependent modifications
of the quark propagator into the deconfined domain, as evident with
$C\not\equiv 1$ in Eq.~(\ref{wsoln}).  It predicts a ``mirroring'' of
finite-$T$ behaviour in the $\mu$-dependence of the bulk thermodynamic
quantities, as illustrated in Fig.~\ref{presfig}.
\begin{figure}[t]
\vspace*{-1.0em}

\centering{\ \epsfig{figure=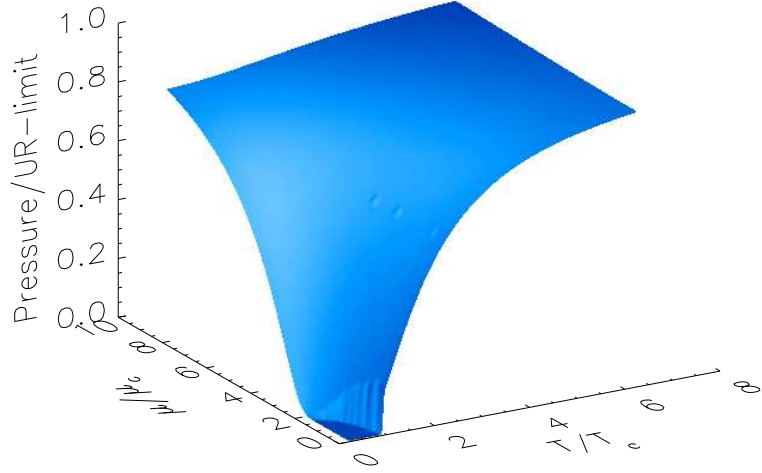,height=7.0cm} }
\caption{The quark pressure, $P_q(\bar T,\bar\mu)$, normalised to the free,
massless (or Ultra-Relativistic) result.\label{presfig}}
\end{figure}

\subsection{$\pi$ and $\rho$ properties}
The model introduced in the last section has also been
used$\,$\cite{schmidt98} to study the $(T,\mu)$-dependence of meson
properties, and to elucidate other features of models that employ a more
sophisticated {\it Ansatz} for the dressed-gluon propagator.  For example,
the vacuum quark condensate is given by the simple expression
\begin{equation}
\label{qbq}
-\langle \bar q q \rangle = 
\eta^3\,\frac{8 N_c}{\pi^2} \bar T\,\sum_{l=0}^{l_{\rm max}}\,
\int_0^{\bar\Lambda_l}\,dy\, y^2\,
{\sf Re}\left( \sqrt{\case{1}{4}- y^2 - \tilde\omega_{l}^2 }\right)\,,
\end{equation}
$\bar T=T/\eta$, $\bar \mu=\mu/\eta$; $l_{max}$ is the largest value of $l$
for which $\bar\omega^2_{l_{\rm max}}\leq (1/4)+\bar\mu^2$ and this also
specifies $\omega_{l_{max}}$, $\bar\Lambda^2 = \bar\omega^2_{l_{\rm
max}}-\bar\omega_l^2$, $\bar p_l = (\vec{y},\bar\omega_l+i\bar\mu)$.  At
$T=0=\mu$, $(-\langle \bar q q \rangle) = \eta^3 /(80\,\pi^2) = (0.11\,
\eta)^3$.
Obvious from Eq.~(\ref{qbq}) is that $(-\langle \bar q q \rangle)$ decreases
continuously to zero with $T$ but {\it increases} with $\mu$, up to a
critical value of $\mu_c(T)$ when it drops discontinuously to zero: behaviour
in agreement with that reported in Secs.~\ref{subsec:D1} and \ref{subsec:D2}.
The vacuum rearrangement emphasised in Sec.~6.4.2 is manifest in the
behaviour of the necessarily-momentum-dependent scalar part of the quark self
energy, $B(\tilde p_k)$ in Eq.~(\ref{ngsoln}).

The leptonic decay constant is also given by a simple expression in the
chiral limit:
\begin{eqnarray}
\label{npialg}
f_\pi^2 & = & \eta^2 \frac{16 N_c }{\pi^2} \bar T\,\sum_{l=0}^{l_{\rm max}}\,
\frac{\bar\Lambda_l^3}{3} \left( 1 + 4 \,\bar\mu^2 - 4 \,\bar\omega_l^2 -
\case{8}{5}\,\bar\Lambda_l^2 \right)\,.
\end{eqnarray}
As anticipated, the combination $\mu^2 - \omega_l^2$, emphasised in
Sec.~6.4.1, characterises the behaviour of Eqs.~(\ref{qbq}) and
(\ref{npialg}).  Without calculation, Eq.~(\ref{npialg}) indicates that
$f_\pi$ will {\it decrease} with $T$ and {\it increase} with $\mu$, which
provides a simple elucidation of the results in Secs.~\ref{subsec:D1} and
\ref{subsec:D2}

The $(T,\mu)$-response of meson masses is determined by the ladder BSE
\begin{equation}
\label{bse}
\Gamma_M(\tilde p_k;\check P_\ell)= - \frac{\eta^2}{4}\,
{\sf Re}\left\{\gamma_\mu\,
S(\tilde p_i +\case{1}{2} \check P_\ell)\,
\Gamma_M(\tilde p_i;\check P_\ell)\,
S(\tilde p_i -\case{1}{2} \check P_\ell)\,\gamma_\mu\right\}\,,
\end{equation}
where $\check P_\ell := (\vec{P},\Omega_\ell)$, with the bound state mass
obtained by considering $\check P_{\ell=0}$.  In this truncation the
$\omega$- and $\rho$-mesons are degenerate.

The equation admits a solution for the $\pi$-meson with
\begin{equation} 
\Gamma_\pi(P_0) = \gamma_5 \left(i \theta_1 
        + \vec{\gamma}\cdot \vec{P} \,\theta_2 \right)
\end{equation}
and the calculated $(T,\mu)$-dependence of the mass is depicted in
Fig.~\ref{pirhomass}.
\begin{figure}
\vspace*{-1.0em}

\centering{\
\epsfig{figure=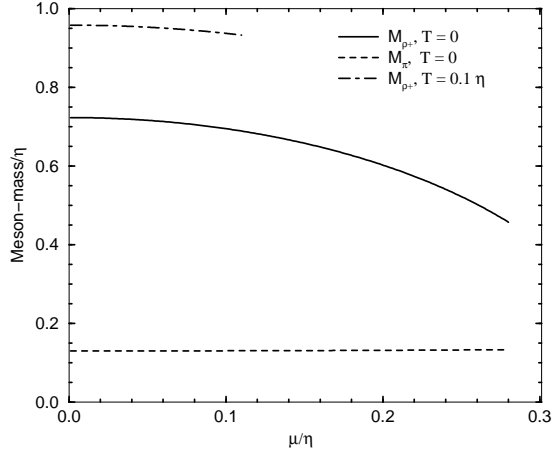,height=7.0cm}}\vspace*{-1.5\baselineskip}
\caption{\label{pirhomass} $M_{\rho+}=M_{\omega+}$ and $M_\pi$ as a function
of $\bar\mu$ for $\bar T = 0, 0.1$.  On the scale of this figure, $M_\pi$ is
insensitive to this variation of $T$.  The current-quark mass is $m=
0.011\,\eta$, which for $\eta=1.06\,$GeV yields $M_{\rho+}= 770\,$MeV and
$M_\pi=140\,$MeV at $T=0=\mu$.}
\end{figure}

For the $\rho$-meson the solution has two components: one longitudinal and
one transverse to $\vec{P}$.  The solution of the BSE has the form
\begin{equation}
\Gamma_\rho = \left\{
\begin{array}{l}
\gamma_4 \,\theta_{\rho+} \\
\left(
\vec{\gamma} - \case{1}{|\vec{P}|^2}\,\vec{P} \vec{\gamma}\cdot\vec{P}\right)\,
        \theta_{\rho-}
\end{array}
\right.\,,
\end{equation}
where $\theta_{\rho+}$ labels the longitudinal and $\theta_{\rho-}$ the
transverse solution.  The eigenvalue equation obtained from Eq.~(\ref{bse})
for the bound state mass, $M_{\rho\pm}$, is
\begin{equation}
\label{rhomass}
\frac{\eta^2}{2}\,{\sf Re}\left\{ \sigma_S(\omega_{0+}^2
        - \case{1}{4} M_{\rho\pm}^2)^2 
- \left[ \pm \,\omega_{0+}^2 - \case{1}{4} M_{\rho\pm}^2\right]
        \sigma_V(\omega_{0+}^2- \case{1}{4} M_{\rho\pm}^2)^2 \right\}
= 1\,.
\end{equation}

The mass of the transverse component is obtained with $[- \omega_{0+}^2 -
(1/4) M_{\rho-}^2]$ in Eq.~(\ref{rhomass}).  Using the chiral-limit
solutions, Eq.~(\ref{ngsoln}), one obtains immediately that
\begin{equation}
M_{\rho-}^2 = \case{1}{2}\,\eta^2,\;\mbox{{\it independent} of $T$ and $\mu$.}
\end{equation}
Even for nonzero current-quark mass, $M_{\rho-}$ changes by less than 1\% as
$T$ and $\mu$ are increased from zero toward their critical values.  Its
insensitivity is consistent with the absence of a constant mass-shift in the
transverse polarisation tensor for a gauge-boson.

For the longitudinal component one obtains in the chiral limit:
\begin{equation}
\label{mplus}
M_{\rho+}^2 = \case{1}{2} \eta^2 - 4 (\mu^2 - \pi^2 T^2)\,.
\end{equation}
The characteristic combination $\mu^2 - \pi^2 T^2$ again indicates the
anticorrelation between the response of $M_{\rho+}$ to $T$ and its response
to $\mu$, and, like a gauge-boson Debye mass, that $M_{\rho+}^2$ rises
linearly with $T^2$ for $\mu=0$.  The $m\neq 0$ solution of
Eq.~(\ref{rhomass}) for the longitudinal component is plotted in
Fig.~\ref{pirhomass}: $M_{\rho+}$ {\it increases} with increasing $T$ and
{\it decreases} as $\mu$ increases.

Equation~(\ref{rhomass}) can also be applied to the $\phi$-meson.  The
transverse component is insensitive to $T$ and $\mu$, and the behaviour of
the longitudinal mass, $M_{\phi+}$, is qualitatively the same as that of the
$\rho$-meson: it increases with $T$ and decreases with $\mu$.  Using $\eta =
1.06\,$GeV, this simple model yields $M_{\phi\pm} = 1.02\,$GeV for $m_s =
180\,$MeV at $T=0=\mu$.

In a 2-flavour, free-quark gas at $T=0$ the baryon number density is $\rho_B=
2 \mu^3/(3 \pi^2)\,$, by which gauge nuclear matter density,
$\rho_0=0.16\,$fm$^{-3}$, corresponds to $\mu= \mu_0 := 260\,$MeV$\,=
0.245\,\eta$.  At this chemical potential the algebraic model yields
\begin{eqnarray}
\label{mrhoa}
M_{\rho+}(\mu_0)  \approx  0.75 M_{\rho+}(\mu=0) &,\; &
M_{\phi+}(\mu_0)  \approx  0.85 M_{\phi+}(\mu=0)\,.
\end{eqnarray}
The study summarised in Sec.~\ref{subsec:D2} indicates that a better
representation of the ultraviolet behaviour of the dressed-gluon propagator
expands the horizontal scale in Fig.~\ref{pirhomass}, with the critical
chemical potential increased by 25\%.  This suggests that a more realistic
estimate is obtained by evaluating the mass at $\mu_0^\prime=0.20\,\eta$,
which yields
\begin{eqnarray}
\label{mrhob}
M_{\rho+}(\mu_0^\prime) \approx  0.85 M_{\rho+}(\mu=0) &,\; &
M_{\phi+}(\mu_0^\prime) \approx  0.90 M_{\phi+}(\mu=0) \,;
\end{eqnarray}
a small, quantitative modification.  The difference between
Eqs.~(\ref{mrhoa}) and (\ref{mrhob}) is a measure of the theoretical
uncertainty in the estimates in each case.  Pursuing this suggestion further,
$\mu=\,^3\!\!\!\surd 2\,\mu_0^\prime$, corresponds to $2\rho_0$, at which
point $M_{\omega+}= M_{\rho+} \approx 0.72\, M_{\rho+}(\mu=0)$ and $M_{\phi+}
\approx 0.85\, M_{\phi+}(\mu=0)$, while at the $T=0$ critical chemical
potential, which corresponds to approximately $3\rho_0$ in
Sec.~\ref{subsec:D2}, $M_{\omega+}= M_{\rho+} \approx 0.65\,
M_{\rho+}(\mu=0)$ and $M_{\phi+} \approx 0.80\, M_{\phi+}(\mu=0)$.  These are
the maximum possible reductions in the meson masses.

This simple model preserves the momentum-dependence of gluon and quark
dressing, which is an important qualitative feature of more sophisticated
studies.  Its simplicity means that many of the consequences of that dressing
can be demonstrated algebraically.  For example, it elucidates the origin of
an anticorrelation, highlighted in Sec.~6.4.1 and found for a range of
quantities, between their response to increasing $T$ and that to increasing
$\mu$.  That makes clear why the transition to a QGP is second order with
increasing $T$ and first order with $\mu$.  Further, in providing an
algebraic explanation of why the $(T,\mu)$-dependence of $(-\langle \bar q
q)\rangle$ and $f_\pi$ is opposite to that observed for $M_{\rho+}$, it
emphasises that the scaling law conjectured in Ref.~[\ref{brownR}] is
inconsistent with many actual calculations that preserve the global
symmetries of QCD.
\section{Some Final Remarks}
There are many challenges in hadronic physics and I have only presented an
outline here.  What should be clear, however, is that this field is
prescribed by the need for nonperturbative methods {\it and}\ models.  The
models are necessary to propagate the qualitatively robust results of
difficult nonperturbative studies into that realm where comparison with the
current and future generation of experiments is possible.

The DSEs serve both purposes: they provide a nonperturbative tool and a model
framework, and have been used widely and efficaciously.  In addition to the
exemplars described herein, there are many other recent studies.  Some are
reviewed in Ref.~[\ref{pctrevR}], others are a calculation of the cross
sections for diffractive, vector meson electroproduction,\cite{pichowsky} the
electric dipole moment of the $\rho$-meson,\cite{martin} and an exploration
of $\eta$-$\eta^\prime$ mixing.$\,$\cite{klabucar}

There are two key phenomenological aspects of contemporary DSE applications:
modelling the infrared behaviour of the gluon propagator and truncating the
kernel of the Bethe-Salpeter equation.  They introduce a model-dependence,
which is restricted to the infrared because the weak-coupling expansion
reproduces perturbation theory and hence the $k^2\gsim 1\,$GeV$^2$ behaviour
is fixed in both cases.  

As always, one should maintain a constructive scepticism about the fidelity
of model input.  Its impact must be gauged; e.g., by exploring and exploiting
the constraints that Ward Identities and Slavnov-Taylor identities can
provide in the theory.  That approach has been particularly fruitful in
QED,\cite{ayse97} and already in the development$\,$\cite{bender96} of a
systematic truncation procedure for the kernel of the BSE in QCD.

Such checks are a useful adjunct.  However, widespread phenomenological
applications remain a key.  Pushing a single framework, successful on a large
domain, to its limits is a very effective means of identifying which elements
of the approach need improvement, and an excellent way to focus resources on
those sites requiring further, fundamental developments.

\section*{Acknowledgments}
I am grateful to the staff of the National Centre for Theoretical Physics at
the Australian National University for their hospitality and support during
the summer school.  This work was supported by the US Department of Energy,
Nuclear Physics Division, under contract no. W-31-109-ENG-38.

\begin{flushleft}

\end{flushleft}

\end{document}